\documentclass{aa}  

\usepackage{CJKutf8} 
\usepackage{graphicx}
\usepackage{txfonts}
\usepackage{lipsum}
\usepackage{subcaption}         
\usepackage{lscape}             
\usepackage{placeins}           
                                
\usepackage[dvipsnames]{xcolor}
\usepackage{url}
\usepackage{siunitx}
\usepackage{longtable}
\usepackage{tabularx}

\usepackage[colorlinks=true, citecolor=blue, linkcolor=blue
]{hyperref}
\usepackage{booktabs}

\newcommand{\mstar}{$M_\star$}

\newcommand{\zh}[1]{\begin{CJK}{UTF8}{bsmi}#1\end{CJK}}

\begin{document}

   \title{Morphology, sizes, and scatter in a large sample of distant quiescent galaxies}



   \author{Gianluca Scarpe\inst{1}\orcid{https://orcid.org/0009-0001-9808-3419}
   \and Francesco Valentino\inst{1,2}\orcid{0000-0001-6477-4011}
   \and Kei Ito\inst{1,2}\orcid{0000-0002-9453-0381}
   \and William M. Baker\inst{3}\orcid{0000-0003-0215-1104}
   \and Antonio Pensabene\inst{1,2}\orcid{0000-0001-9815-4953}
   \and Pengpei Zhu (\zh{朱芃佩}) \inst{1,2,4}\orcid{0000-0002-6768-8335} 
    }

   \institute{DTU Space, Technical University of Denmark, Elektrovej 327, DK- 2800 Kgs. Lyngby, Denmark
    \and 
        Cosmic Dawn Center (DAWN), Denmark
    \and
        DARK, Niels Bohr Institute, University of Copenhagen, Jagtvej 155A, DK-2200 Copenhagen, Denmark
    \and 
    INAF-Osservatorio Astrofisico di Arcetri, Largo Enrico Fermi 5, I-50125 Firenze, Italy}
       
    \date{Received month day, year}
    \titlerunning{}
    \authorrunning{Scarpe et al.}
 
  \abstract
   {Massive galaxies stopped forming stars surprisingly early in the history of the Universe. However, after quenching galaxies keep growing in size across time, as established in the literature up to cosmic noon. In this work, we assemble one of the largest and most comprehensive multi-wavelength photometric sample of massive quenched galaxies at $z>3$ from publicly available JWST observations, counting 137 quiescent candidates within $\sim 825\:\mathrm{arcmin^2}$ and redshift $3 \leq z < 7$ across the well-studied extragalactic fields GOODS, PRIMER, and CEERS. We model their surface brightness distribution across 5 bands mapping their UV-to-NIR rest-frame wavelength with Sérsic profiles, and derive their sizes, concentrations, and ellipticities. The size-mass relation is consistent with previous studies, showing a shallower slope at low redshift $z \in [3, 3.5]$ and a steeper slope at high redshift $z > 3.5$. Size decreases with increasing redshift, in agreement with previous studies and we extend them up to $z_\mathrm{spec} = 4.9$. Leveraging the large sample statistics, we robustly constrain the intrinsic scatter of the mass--size relation to $\sim 0.3\:\mathrm{dex}$ with no relevant dependence on the redshift and the filter used. 
   At the population level, our multi-wavelength modeling reveals that size decreases with increasing observed wavelength, and the wavelength gradient decreases with increasing stellar mass. This result proves that in the NIR-bands massive elliptical galaxies appear more compact. The Sérsic index does not show significant dependence on wavelength, independently of the stellar mass. Following a forward--backward Bayesian fit analysis to assess the significance of several parameters in predicting the size of our sample, we do not identify any significant secondary dependence of the size on axis ratio $q$, S\'ersic index $n$, $UVJ$-colors, and environment. The combination of stellar mass and redshift is sufficient to predict the size of quiescent galaxies at $z>3$, albeit with a large scatter. This suggests that the commonly used parameters of a Sérsic distribution cannot explain the large intrinsic scatter around the stellar mass--size relation, suggesting that other physical quantities need to be taken into account to break the degeneracy between evolution paths across the galaxy population.}


   \maketitle
\nolinenumbers

\section{Introduction}

Quiescent galaxies (QGs) are fascinating and widely studied systems that show little -- if any -- star formation compared to galaxies of similar mass observed at the same cosmic epoch. They were initially identified in the local Universe in connection with studies of elliptical galaxies, i.e. ``early-type'', spheroidal systems that are nearly featureless when compared to spiral or, more generally, late-type galaxies \citep{tuning_fork_1996}. Thanks to the improved sensitivity of astronomical instruments, most recently \textit{James Webb} Space Telescope (JWST), which enables the detection of near-infrared (NIR) starlight emission at high redshift, interest in these objects has grown substantially, with quiescent galaxies now spectroscopically confirmed out to $z \simeq 4$–$7$ \citep{degraaff_2024, carnall2024, Weibel_2025, baker2025d, Ito2025}. In the classical formation paradigm, the high stellar surface densities typically observed at high redshift are interpreted as the result of rapid, intense star formation occurring over short timescales, followed by an abrupt quenching phase \citep{cimatti_2008, toft_2014, Lagos2025, deLucia_2025}. Among the most widely accepted mechanisms responsible for halting star formation in massive galaxies and on short timescale is active galactic nucleus (AGN) feedback, which is thought to play a key role by expelling gas and heating the surrounding material and prevent gas accretion, thereby suppressing further star formation \citep[e.g.,][]{Benson_2003, Bower_2006, hopkins_2006, Kurinchi-Vendhan_2024, Lagos2025, Farcy_2025}.

After quenching, galaxies continue to grow despite the absence of ongoing star formation \citep{daddi_2005, Trujillo_2006, Toft_2007, Cassata_2013, vanderwel}, implying that additional physical mechanisms drive their subsequent evolution. This growth could also coincide with and contribute to morphological transformations (e.g., from disky to spheroidal systems). Consequently, galaxy morphology, namely the shape and the associated light distribution, provides a powerful tool for tracing evolutionary pathways before and after quenching. Studies have shown that the size of galaxies, inferred as the half-light radius or semi-major axis, correlates with stellar mass and formation history, including the mechanisms through which galaxies quenched and subsequently evolved \citep{bezanson_2009}, revealing a diverse set of evolutionary channels from quenching to post-quenching growth. Mergers are thought to play a key role in both quenching and the later evolution of galaxies \citep{bezanson_2009, Ito_2025_merger}. Different merger channels have been proposed, including (dry) major mergers, (dry) minor mergers \citep{Naab_2009}, and mini-mergers \citep{Naab_2009, Oser_2012, Suess_2023, Nipoti2025}, the latter being able to puff up sizes without dramatically affecting the total mass of the massive progenitor. 

Numerous studies have investigated galaxy morphology using a variety of techniques and levels of detail, consistently finding an increase in galaxy size with stellar mass \citep{shen2003, Cassata_2013, vanderwel, Kubo_2018, Mowla_2019, Lustig_2020, Kawinwanichakij_2021, nedkova21, Forrest_2022, Ito_2024, Wright_2024, Ji_2024, genin2025, kawinwanichakij2025, allen2025, Yang, Trujillo_2006, chen2026}. At fixed stellar mass, galaxy size increases with cosmic time \citep{vanderwel, Ito_2024}. However, this size evolution may be affected by the so-called ``progenitor bias'' \citep{Barro_2013, Carollo_2013}, whereby newly quenched galaxies, typically having sizes similar to those of star forming galaxies (SFGs), enter the quiescent population at later times, leading to an overestimation of the intrinsic size growth of individual galaxies. Furthermore, quiescence is not always associated with a bulge-dominated morphology. Some QGs, especially at high redshift, exhibit disky structures and possibly rotation-supported stellar dynamics, suggesting that the quenching of star formation precedes the morphological transformation into early-type galaxies \citep{Toft_2017, Slob_2025}. Observationally, measured galaxy sizes also depend on wavelength. To mitigate this effect, several studies have adopted a normalized optical rest-frame wavelength ($0.5\,\si{\mu m}$, \citealt{nedkova21, Kawinwanichakij_2021}), but only the advent of \textit{JWST} has enabled systematic mapping of rest-frame optical emission at $z>3$ \citep{Ito_2024}.\\ 

It is therefore evident that accurate morphological and size measurements of galaxies observed close to their quenching epoch provide a stringent test for galaxy formation and quenching models. In this paper, we assess the morphologies and sizes of quiescent systems at $z>3$ by exploiting several years of multi-wavelength \textit{JWST}/NIRCam observations across some of the best-studied extragalactic fields distributed over the sky. 
Our goal is to provide the most up-to-date and statistically significant, purely observational benchmark for morphological studies of high-redshift quiescent galaxies. We present a morphological analysis of $\sim130$ quiescent galaxies at $z\geq3$, corresponding to a sample that is $\sim5\times$ larger than those used in previous studies (Section \ref{selection}). By modeling the stellar light profile in five bands blueward and redward of the Balmer break characteristic of quiescent galaxies, in Section \ref{size_computation} we investigate the presence of wavelength-dependent gradients in galaxy size and Sérsic index, along with their dependence on stellar mass. We then derive the stellar mass–size relation for quiescent galaxies at $z=3$–$7$ and, crucially, determine its intrinsic scatter -- an analysis made feasible by our large statistical sample. We further explore the incidence of disks embedded within central bulges. Finally, in Section \ref{result_analysis}, we attempt to identify the optimal linear combination of primary morphological and photometric parameters for predicting the sizes of high-redshift quiescent galaxies, and we investigate the possible drivers of the large intrinsic scatter in the mass–size relation. Throughout this paper, we assume a $\Lambda$CDM cosmology with $H_0 = 70\,\si{km\:s^{-1}Mpc^{-1}}$, $\Omega_m = 0.3$, and $\Omega_\Lambda = 0.7$. Magnitudes are given in the AB system, and a \citet{Chabrier_2003} initial mass function (IMF) is adopted.

\section{Data}
\label{data}
In order to build a meaningful statistical sample, we considered data from major extragalactic fields covered by JWST imaging observations: PRIMER-UDS (266.7 $\mathrm{arcmin^2}$), PRIMER-COSMOS (260.1 $\mathrm{arcmin^2}$, \citealt{Dunlop2021, donnan2024}), the Cosmic Evolution Early Release Science Survey (CEERS, 109.5 $\mathrm{arcmin^2}$, \citealt{Finkelstein_2025}), the GOODS-S (70.5 $\mathrm{arcmin^2}$), and GOODS-N (120.9 $\mathrm{arcmin^2}$) fields \citep{giavalisco2004, grogin2011, koekemoer2011}. 
These fields cover a cumulative area $\approx 825\:\mathrm{arcmin}^2$. We analyzed photometric data collected with the broad-band filters F150W, F200W, F277W, F356W, and F444W from the DAWN JWST Archive\footnote{\url{https://dawn-cph.github.io/dja/blog/2024/08/16/morphological-data/}} (DJA, \citealt{valentino2023ApJ}). For the color selection described in the next section, we also included observations with JWST/MIRI and with the \textit{Hubble} Space Telescope (HST), when available. The photometry was extracted within circular apertures with a diameter of $0\farcs5$ and corrected to total fluxes based on the detection image, which is a combination of the available long-wavelength NIRCam filters. In the v7 version of the available mosaics, all images in the CEERS and PRIMER fields are drizzled to a common pixel scale of $0\farcs04$ across all bands. The GOODS-S and GOODS-N images obtained with the F150W and F200W filters have a pixel size of $0\farcs02$, while each pixel in the remaining LW filters corresponds to $0\farcs04$. For the surface brightness modeling, we used the point-spread functions (PSFs) computed by \citealt{genin2025} and available in the DJA. Each PSF has the same pixel size as the reference image.

\section{UVJ-color selection}
\label{selection}
\begin{figure}
    \centering
    \includegraphics[width=1\linewidth]{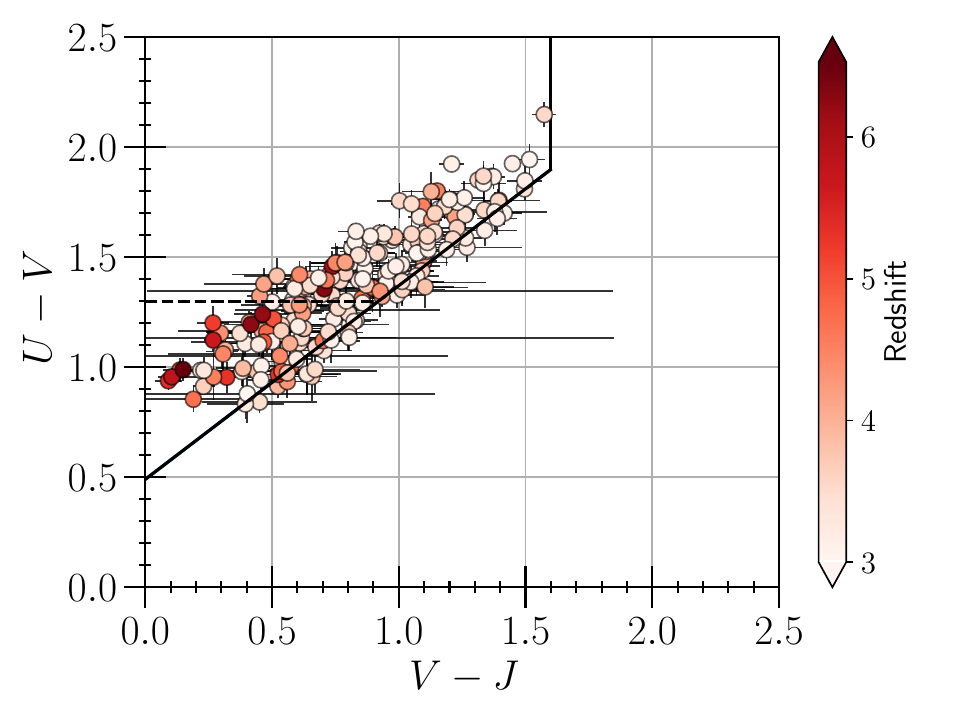}
    \caption{$UVJ$ diagram of the final vetted sample after visual inspection, color coded by redshift. The black lines show the selection box in \citet{williams2009}, extended at the blue end to include recently quenched systems \citep{Schreiber_2018, Belli2019}.} 
    \label{fig:goodSelUVJ}
\end{figure}

We selected quiescent galaxies (QGs) based on their $UVJ$ rest-frame colors \citep{williams2009}. We adopted the colors and their uncertainties computed with \texttt{eazy-py} \citep{brammer2008}, and publicly available on DJA. We adopted the  criteria presented in  \citealt{Schreiber2015} and \citealt{Belli2019}:
\begin{align}
    U-V &\geq 1.3\\
    V-J &\leq 1.6\\
    U-V &\geq 0.88(V-J)+0.49
\end{align}
allowing for the $1\sigma$ uncertainty on the rest-frame colors. 
 Here we focus on $3 \leq z < 7$ and $M_\star\geq10^{9}\,M_\odot$ to select distant, classically selected, bona fide quenched objects and avoid contamination of blue, low-mass galaxies in temporary phases of quiescence due to the burstiness of their star formation history (SFH) \citep{strait_2023, looser_2024, Baker_2025_remnant}. The mass cut is well above the 90\%\ stellar mass completeness limit of $\approx 10^7\:M_\odot$ in all fields. 
Photometric redshifts and stellar masses in our selection were also computed with \texttt{eazy-py} on the JWST and HST photometry on the aperture-corrected photometry as part of the catalogs available on DJA, adopting a \citet{Chabrier_2003} IMF, stellar templates from the Flexible Stellar Populations Synthesis code \citep[FSPS;][]{conroy_2010}, the \cite{kriek_2013} dust attenuation law, and allowing for emission lines on pre-computed CLOUDY grid \citep{Byler_2018}. More details on this can be found in the literature \citep{Gould_2023, valentino2023ApJ}.
We further set a limit on $\chi_{\mathrm{raw}}^2 < 50$, where $\chi_{\mathrm{raw}}^2$ is used to assess the quality of the fit performed with stellar templates in the \texttt{eazy-py} models, to remove galaxies with poorly constrained spectral energy distributions (SEDs). 
To remove faint galaxies, typically characterized by shallow Balmer breaks and blue slopes in the rest-frame UV, we applied a S/N cut of $S/N > 150$ in $0\farcs5$ apertures in the LW-combined detection image, resulting in $1172$ objects -- of which $83\%$ have $M_\star<10^{10.3}M_\odot$ (low-mass), the mass limit we will impose later to derive meaningful comparisons with the existing literature. The threshold in signal-to-noise ratio is applied to ensure excellent quality on the morphological fitting and allowing for sampling scales smaller than the PSF \citep{Toft_2007}. A visual inspection of the SEDs conducted by two members of the team (GS, FV) ensured their quality, resulting in a final, vetted sample of $137$ objects, $25\%$ of which have $M_\star<10^{10.3}M_\odot$ while still displaying meaningful Balmer breaks typical of recently quenched systems (Figure \ref{fig:goodSelUVJ}). Sources with uncertain SEDs and poorly constrained photometric redshift solutions ($\sigma_{\rm z_{phot}}>1$), dusty contaminants, and objects with shallow Balmer breaks and steep bright blue upturns were excluded at this stage. We note that, without the pre-adoption of the S/N cut, necessary to ensure an excellent morphological analysis, there would be 36 more galaxies in our sample. 

In general, a pure $UVJ$ color selection is affected by contamination from dusty SFGs up to a 20\% level \citep{Schreiber_2018}. We anticipate that the results in the rest of the paper are not significantly affected for a random distribution of the contaminants in the mass and redshift ranges covered by our final sample (Section \ref{MCMC fitting size}). We also note that the inclusion of MIRI photometry, when available, allows in principle for the derivation of robust stellar masses and the possible detection of very distant sources \citep{Wang_2025_miri, Yang_2026}. Moreover, the cut in S/N in the LW filters does not prevent us from selecting fainter QGs in the bluest bands, possibly contributing to the ``NIR-faint'' populations selected pre-JWST \citep{Wang_2019}. As an example, we recover all the spectroscopically confirmed QGs in the follow-up of NIR-faint sources in \cite{Barrufet_2025}, which fulfill our mass and S/N criteria.\\
For full transparency and reproducibility of the visual inspection, the photometric SEDs and best-fit {\tt eazy-py} models of the final sample and the excluded sources are available online.

\subsection{Comparison with recent samples of quenched galaxies}

Before proceeding with the morphological analysis, we put our selection of distant quenched galaxies in the context of recent literature on this topic. Our color criteria select only the oldest and most massive quenched galaxies in \cite{baker2025b} sample, also based on DJA products in the same fields, since we do not consider the same extension towards bluer colors and higher specific star formation rate (sSFR) sources. To quantify this difference, we calculate the comoving number densities in four redshift bins between $z=3-7$ based on the total area covered by the observations, and the uncertainties are computed by taking the 84th-16th percentile on the redshift bin (Table \ref{tab:density}). Our estimates show lower comoving number densities than \citet{baker2025b}. With a mass cut at $\log (M_\star/M_\odot) \ge 10.48$, corresponding to the mass limit in \citet{Nanayakkara_2025}, which inherited a selection similar to ours presented in \citealt{Schreiber_2018}, we find a slightly lower number density.
 
Finally, we find an overlap with 12/17 of the sources in the fields considered here and at $z \geq 3 $ in the spectroscopic compilation based on JWST/NIRSpec data in \citet{Ito2025}. The mismatch is due to the extended selection they used, which included $D_n4000$ and cuts on sSFR, in contrast to our pure initial-color selection. The comparison between the photometric redshifts considered here and the spectroscopic values in \citealt{Ito2025} excludes only one source with $z_{\rm spec}<3$, thus highlighting the robustness of the photometric solutions.


\begin{table}[h]
\centering
\caption{Comoving number densities.}
\begin{tabular}{ccc}
\toprule
\toprule
$z$ & $n_{\rm comoving}$ & $N_{\rm objects}$\\
 & \small $[10^{-5} \times \si{Mpc^{-3}}]$ & \\
\midrule
$3.0-3.5$ & $5.2^{+1.0}_{-2.6}$ & 76 \\[0.5ex]
$3.5 - 4.0$ & $2.7^{+0.7}_{-1.5}$ & 38\\[0.5ex]
$4.0 - 5.0$ & $0.8^{+0.2}_{-0.4}$ & 20 \\[0.5ex]
$5.0 - 7.0$ & $0.07^{+0.02}_{-0.02}$ & 3 \\ [0.5ex]
\bottomrule
\end{tabular}
\tablefoot{The uncertainties on the comoving number densities do not include the effect of the cosmic variance. Nevertheless, its impact amount to $\sim15$\% at $z\geq3$ for such a sample in the combined fields considered in this work \citep{baker2025b}.}
\label{tab:density}
\end{table}

\begin{figure*}
    \centering
\includegraphics[width=1\linewidth]{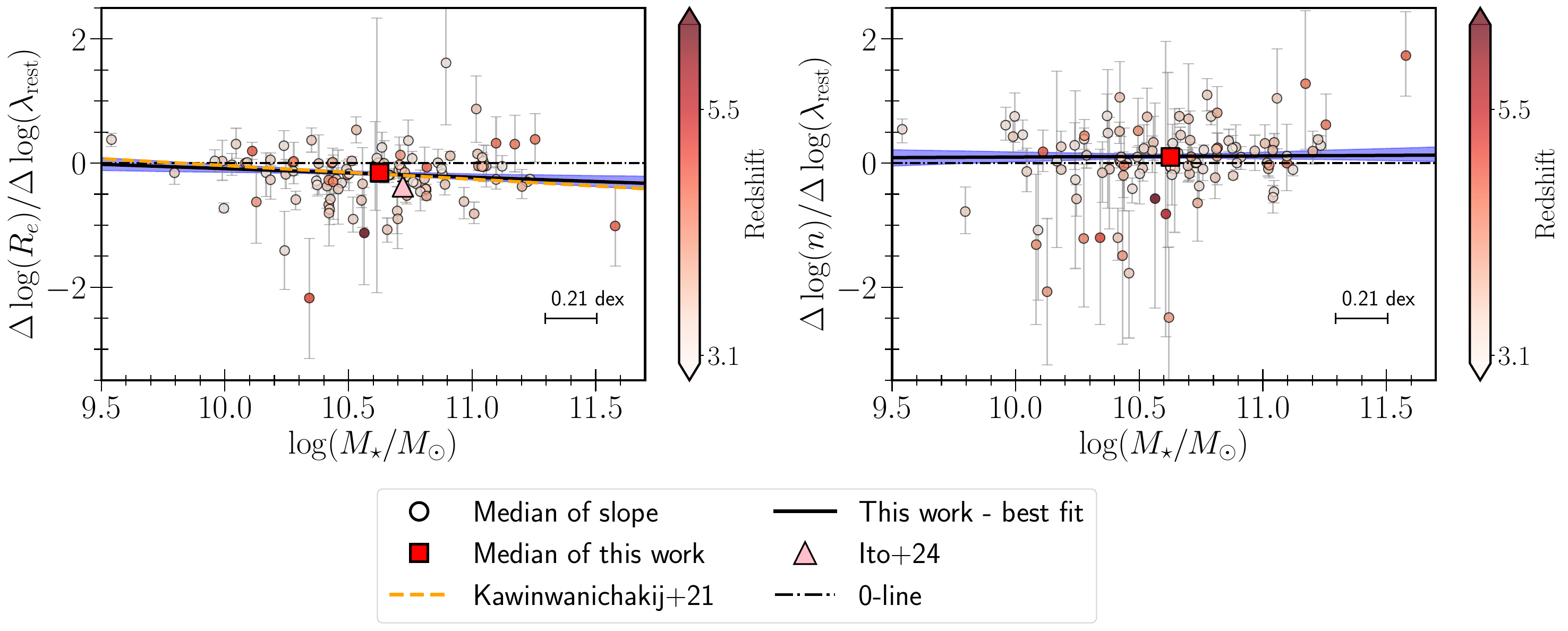}
    \caption{Wavelength gradients of the effective size and S\'ersic index as a function of the stellar mass. Our sample is indicated by filled circles, color-coded by redshift. The uncertainty on the stellar mass is fixed at $0.21$ dex. 
    The median value of the sample is marked by the red filled square, while the pink triangle shows the result in \citealt{Ito_2024}. The black solid line and blue shaded area indicate the best-fit model and its uncertainty. The dashed orange line in the left panel indicates the relation in \citet{Kawinwanichakij_2021}. The dashed dotted lines indicate a constant gradient as a function of stellar mass.}
    \label{fig:slopes_double}
\end{figure*}

\section{Morphological analysis and size estimates}
\label{size_computation}
To leverage the quality and depth of the NIRCam data, we conducted a multi-wavelength morphological analysis and surface brightness modeling. Previous studies \citep{Ito_2024, vanderwel} suggest gradients in galaxy sizes with wavelength, which we investigate in our sample. We then derive the stellar mass--size relation and its intrinsic scatter, now accessible thanks to the large available statistics. Finally, we present two-component models and evaluate their ability to capture structures missed by simple S\'{e}rsic profiles.

\subsection{Surface brightness modeling}
\label{pysersic}
Our primary analysis relies on the results derived from fitting the surface brightness of galaxies in our sample using a simple S\'ersic \citep{Sersic_1963} profile with the Bayesian code \texttt{pysersic} \citep{pysersic2023}.
We generated cutouts of science and weight images, including Poissonian noise, of $2\farcs0\times2\farcs0$ ($\sim 15\,\mathrm{kpc} \times 15\, \mathrm{kpc}$ at $z=3$) centered around each galaxy.
We did not mask potential contaminants in the cutout area and modeled with a multi-source Sérsic profile all the observed sources. The prior was estimated from data retrieved with \texttt{SEP} \citep{SExtractor,Barbary_2016} and the sky was automatically estimated with the function \texttt{estimate\_sky} of \texttt{pysersic}.
However, in 11 cases, we simultaneously remodeled close companions for which the automatic fit failed by forcing the detection of multiple sources (in F444W). Unsuccessful deblending of 9 galaxies with extended features motivated us to remove those sources from the rest of the morphological analysis. The final sample consists of 129 objects\footnote{Consider that different filters have different number of sources. We match the objects in F444W to F200W, since 8 sources are not detected in the latter filter. The total number of sources we refer to is in F200W because it is the one with the lowest number of detections.}.\\ 

For the full multi-wavelength analysis, we considered models spanning the whole set of filters, which included the F150W, F200W, F277W, F356W and F444W filters.
We thus opted for a simple IB approach, which can accommodate for fitting images in all fields irrespectively of their different pixel sizes (Section \ref{data}), and that provides much flexibility for excluding ill-detected features in short wavelength filters. 

\subsection{Morphological gradients with wavelength}
\label{multiband_section}

For the following calculations, we include sources within the 80\% percentile of the sizes and S\'{e}rsic indices distributions in F150W, which probes the blue end of the Balmer break at $z>3$. This is to ensure that outliers in the effective semi-major axis ($R_{e}$ or ``size'' hereafter) and the S\'ersic index ($n$) estimates due to the intrinsically faint emission at these wavelengths do not bias the final model. 
We quantify the gradients of each galaxy using a linear model \citep{vanderwel, Ito_2024, kawinwanichakij2025, Ji_2024}:
\begin{align}
    \log \left[ \theta(\lambda_\mathrm{rest})\right] = \gamma_\theta\log\left(\frac{\lambda_\mathrm{rest}}{0.5\:\si{\mu m}}\right) + \log \left[ \theta(0.5\:\si{\mu m})\right] 
\end{align}
where $\theta$ is $R_e$ or $n$, and $0.5\:\mathrm{\mu m}$ is the optical rest-frame normalization wavelength. The mean gradients of the sizes and S\'ersic indices with wavelengths are $\gamma_{R_e} = -0.19\pm 0.03$ and $\gamma_{n} = 0.01\pm 0.04$, respectively. For the sample we find a distribution of medians $\{\gamma_{\theta,i} \}_{i=1}^{103}$. The median and 16-84\% ranges of the distribution are $\gamma_{R_e} = -0.15^{+0.26}_{-0.43}$ and $\gamma_{n} = 0.10^{+0.41}_{- 0.54}$. We thus find evidence for the existence of slightly negative size gradients with wavelengths. On average, we do not find significant gradients of the S\'{e}rsic index with wavelengths in our sample of distant QGs -- yet, the large scatter indicates substantial diversity in the behavior of individual objects. Recent literature suggests that Sérsic index does increase with mass for $ z \leq 2.5$ \citep[][]{martorano2025} and previous works \citep{vanderwel, Ito_2024, Kawinwanichakij_2021} also suggested the existence of mild systematic variations of gradients with stellar mass for QGs at lower redshifts. We can now test these results with our statistical sample at $z>3$. Assuming that $\gamma_\theta = \gamma_\theta(M_\star)$, we fit the following relation:
\begin{align}
    \gamma_\theta(M_\star) = m_\theta\log\left(\frac{M_\star}{M_\odot}\right)+q_\theta
\end{align}
The dependence of the size gradient $\gamma_{Re}$ on the stellar mass shows a minimal negative slope though not consistent with a constant flat relation within $1\sigma$ (Figure \ref{fig:slopes_double}), in agreement with previous results at lower redshifts \citep{Kawinwanichakij_2021}. The multi-band modeling returns effective negative gradients of effective radii with wavelength for 70/103 galaxies and 42/103 for S\'{e}rsic indices. The size gradient, in particular, points out to a population in which inside-out quenching is the dominant feature. Furthermore, we find a very weak dependence on stellar mass for $\gamma_n$, such that more massive galaxies tend to have more positive gradients (higher $n$ at longer wavelengths). The resulting mass trends are parameterized as:
\begin{align}
    \gamma_{R_e}(M_\star) &= -0.14^{+0.08}_{-0.08}\log\left(\frac{M_\star}{M_\odot}\right)+1.29^{+0.86}_{-0.87}\\
    \gamma_n(M_\star) &= 0.02^{+0.11}_{-0.11}\log\left(\frac{M_\star}{M_\odot}\right)-0.13^{+1.21}_{-1.21}
\end{align}
and \texttt{emcee} \citep{Foreman-Mackey_2013} yields the intrinsic scatters $\sigma(\gamma_{R_e}) = 0.25^{+0.03}_{-0.03}$, $\sigma(\gamma_n) = 0.26^{+0.04}_{-0.04}$ respectively. Our removal of the 20\% of the objects (see above) does not significantly impact the final conclusions, by only giving a $25\%$ steeper gradients but may underestimate the intrinsic scatter. At fixed mass $M_\star = 5\times10^{10}M_\odot$ the results are consistent with the average trends. It turns out that the effective radius is weakly affected by the filter used, showing a negative mean correlation. The mean values of the size gradient is not consistent with \citealt{Ito_2024} who presents $\gamma_\mathrm{R_e}=-0.38\pm 0.08$ although with slightly larger median mass (Figure \ref{fig:slopes_double}). Notwithstanding, mean values are prone to selection bias and the deviation from the mean only tells how well the mean is predicted.
On the population level, the Sérsic index has no discernible trend in wavelength and mass. Finally, the results hereby obtained indicate that at longer wavelength more massive galaxies are mapped to more compact sources and with steeper gradient.

\begin{figure*}
    \centering
    \includegraphics[width=\textwidth]{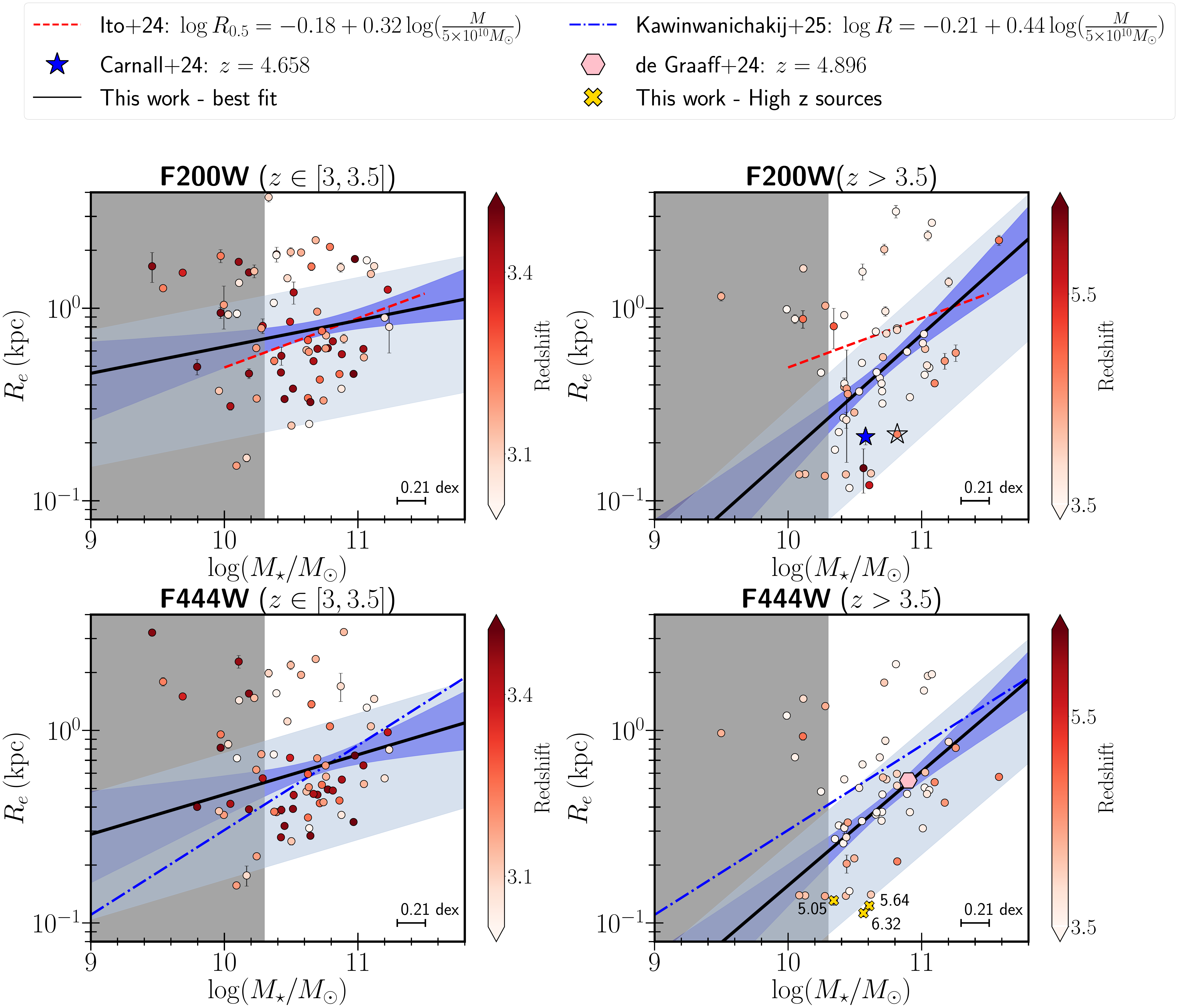}
    \caption{Stellar mass–size relation for quiescent systems at $z>3$. Red filled circles, color-coded by redshift, represent our measurements. Size uncertainties correspond to the 16–84\% interpercentile range, while stellar mass uncertainties are fixed at $0.21$ dex and are indicated by black segments in the lower-right corner of each panel. In each panel, solid black lines and dark blue shaded regions show the best-fit relation and its uncertainty extracted as the 16–84\% interpercentile range from the \texttt{emcee} chain, respectively, while the light blue symmetric shaded region indicates the intrinsic scatter of the relation. The gray area marks the interval of mass that is not used for the fit. The relations from \citet{Ito_2024} and \citet{kawinwanichakij2025} are shown as red dashed and blue dash–dotted lines, respectively. We highlight individual sources as follows: the original mass and size measurements of the quiescent galaxy at $z=4.6$ from \citet{carnall2024} are marked with a blue star, while our measurements for the same object are shown with a black open star; the galaxy at $z=4.9$ from \citet{degraaff_2024} is indicated by a pink hexagon; for reference, our quiescent galaxy candidates at $z>5$ are marked by golden crosses in the bottom-right panel. The filter and redshift bin corresponding to each panel are indicated at the top.}
    \label{fig:F200+F444_split}
\end{figure*}

\subsection{The stellar mass-size relation at \texorpdfstring{$z\ge3$}{TEXT}}
\label{MCMC fitting size}
The relation between $R_e$ and $M_\star$ in QGs has already been probed and successfully established numerous times for $z \lesssim 3$ in the literature. The extension to $z\sim3-5$ featured remarkable results and has already yielded deep insights into the analytical relation connecting these two quantities \citep{vanderwel, Ito_2024, kawinwanichakij2025,allen2025, baker2025c,Ji_2024, chen2026}. However, the number of quiescent galaxies used to derive this relation at these redshifts in previous studies had been limited ($\sim20-30$). We thus reexamine this relation by using our $\sim 5\times$ larger statistical sample.\\ 

For the sake of simplicity, here we focus on measurements in the F200W and F444W bands to provide a reference for the optical and near-infrared rest-frame sizes, keeping in mind our findings about mild gradients with wavelengths in the previous section.
For the stellar masses, we adopt the estimates derived with \texttt{eazy-py} publicly available on DJA. These stellar masses are overall consistent with values from classical template fitting and their uncertainties amount to $0.21$ dex \citep{Gould_2023, Ito_2024}, a contribution to the error budget that we fold in the modeling of the mass--size relation\footnote{See \href{https://zenodo.org/records/18669101?preview=1&token=eyJhbGciOiJIUzUxMiJ9.eyJpZCI6ImMzYmNlODUzLWQwYTktNDdiYi05ZDQyLTBiMWU1YzI5YjJlOSIsImRhdGEiOnt9LCJyYW5kb20iOiJkNDM4YjkzNzY4ZWZiODM3NDE4MmU3OGI2MDNmMjMxOSJ9.IVMoA4i5csQY9J1VSCZTYVXtkbNN1kredIkizICWpxqamlRAXkk4Y69LNf60mjjIHZ8KeEgEbEyGbyC0lFAXcA}{Zenodo} for the formulas.}.
By following a common approach in the literature \citep{vanderwel}, we model the relation as: 
\begin{align}\label{sizemassdep}
    \log \left(\frac{R_e}{\si{kpc}} \right) = \log \left(A\right)  + \alpha \log \left(\frac{M_\star}{5\times 10^{10}\,M_\odot}\right)
\end{align}

\begin{figure}
     \centering
         \centering
         \includegraphics[width=\columnwidth]{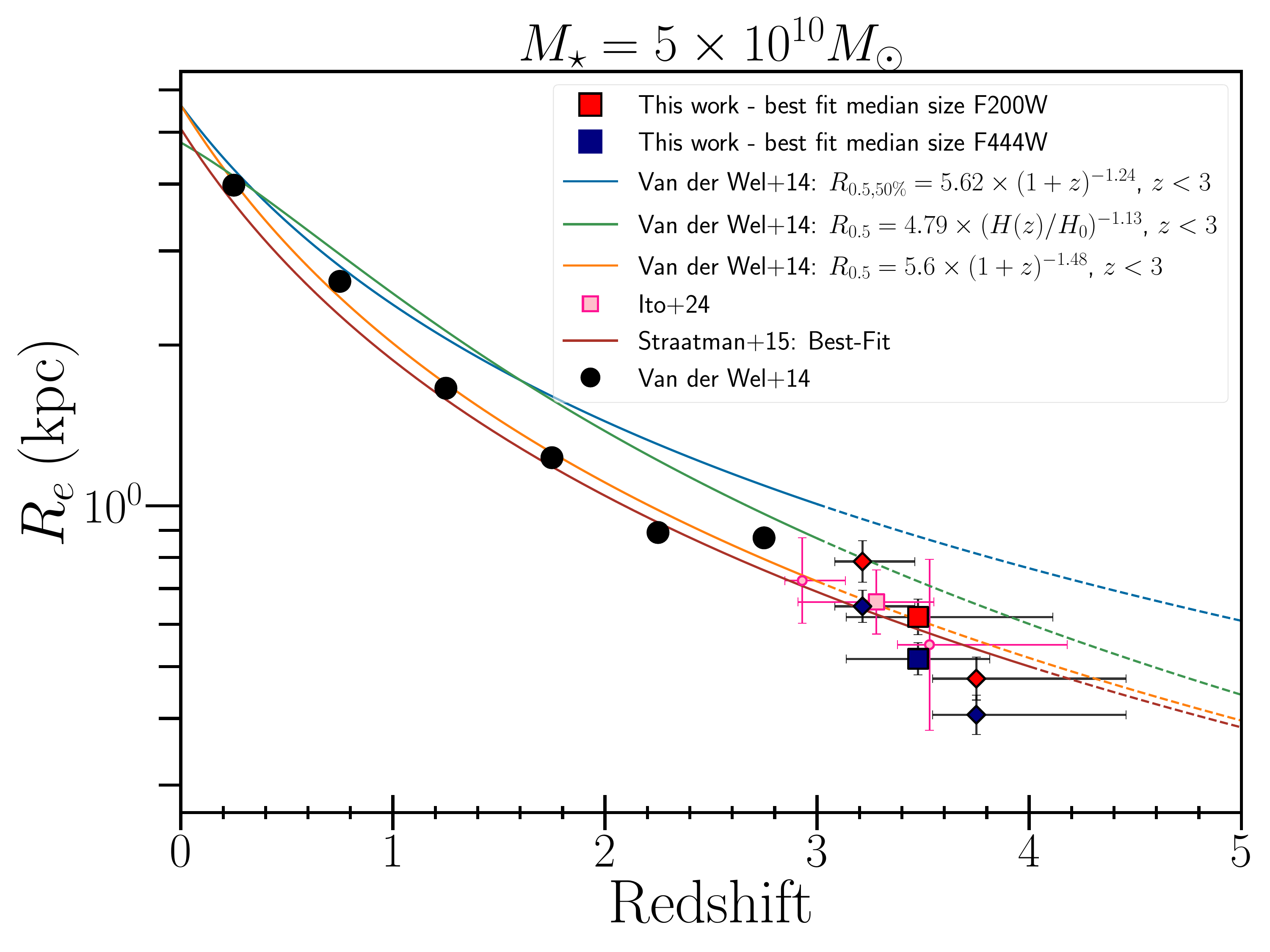}
        \caption{Median size evolution for a $M_\star=5\times10^{10}\,M_\odot$ quiescent galaxy with redshift. The red and blue symbols indicate our median measurements in the F200W and F444W filters, respectively; squares show the median of the sample in the $z=3-5$ range, while diamonds show binned samples. Pink square and circles show the equivalent medians from \citet{Ito_2024}. The black circles indicate the measurements in \citet{vanderwel}. Best-fit models in the literature \citep{vanderwel, Straatman_2015} and their extrapolations beyond their redshift of validity are shown as solid and dashed lines.
        } 
       \label{fig:z_dep_all}
\end{figure}

We define a  likelihood  assuming that the radius $R_e$ follows a log-normal distribution $\mathcal{N}(\log R_e, \delta \log R_e)$, where $\delta \log R_e$ is the total error on the observed radius, namely including the statistical error from \texttt{pysersic} and the projected mass uncertainty \citep{vanderwel, shen2003}. If the expected distribution has expectation value $\log \hat{R}_e = \log R_e (\hat{\alpha})$ and intrinsic scatter $\sigma_{\log R_e}$, the likelihood will be obtained as the convolution of two log-normal distributions (see \href{https://zenodo.org/records/18669101?preview=1&token=eyJhbGciOiJIUzUxMiJ9.eyJpZCI6ImMzYmNlODUzLWQwYTktNDdiYi05ZDQyLTBiMWU1YzI5YjJlOSIsImRhdGEiOnt9LCJyYW5kb20iOiJkNDM4YjkzNzY4ZWZiODM3NDE4MmU3OGI2MDNmMjMxOSJ9.IVMoA4i5csQY9J1VSCZTYVXtkbNN1kredIkizICWpxqamlRAXkk4Y69LNf60mjjIHZ8KeEgEbEyGbyC0lFAXcA}{Zenodo}).
Moreover, low-mass galaxies have a higher comoving number density than high-mass galaxies \citep[][for recent stellar mass functions of QGs at high redshifts]{baker2025b, Shuntov_2025_SMF}.  Therefore, we introduce a weight in our model to compensate for the bias between galaxies of different mass \citep[e.g., ][]{vanderwel}. The weight is selected to be the inverse of the Stellar Mass Function (SMF) $\Phi(M_\star)$.  For a direct comparison with the recent results at $z=3-4$ in \citealt{kawinwanichakij2025}, we used their parameters for the SMF: 
\begin{align}
    \Phi(M_\star) = \left(\frac{M_\star}{M_0}\right)^{-0.41}\exp\left(-\frac{M_\star}{M_0}\right)
\end{align}
with $M_0=10^{10.41}\,M_\odot$.
The parameters of the SMF are computed for $z \in (3, 3.5)$ in \citet{Weaver_2023} and they have already been widely used in the literature. The  interval $z \in (3, 3.5)$ approximately corresponds to $1/5$ of the whole range spanned by our galaxies, but also encloses 50\% of the sample.
The log-likelihood becomes $\mathcal{L} = \ln(f\cdot\Phi^{-1})$, $f$ being the non-weighted likelihood, and for $N$ samples:
\begin{align}
    \mathcal{L} = \sum_{i = 1}^N \ln \left[f_i\cdot\Phi(M_{i, \star})^{-1}\right]
\end{align}
The prior for the modeling is chosen in accordance to \citealt{kawinwanichakij2025}: $\alpha \in (0, 1)$, $\log A \in (-2, 1)$, $\sigma_{\log R_e} \in (0.05, 0.5)$, choosing 50 walkers and 10000 steps like in \citet{nedkova21}. Nevertheless, we note that accounting for the weighting with $\Phi$ does not appreciably change our results. Wishing to maintain a high consistency with previous literature \citep{vanderwel, Ito_2024}, only objects with $\log \left(M_\star/M_\odot \right) > 10.3$ were included in the fit. This is also consistent with the high-mass sample in the recent work by \cite{chen2026}. As these authors point out, a bending of the relation is present below approximately this mass threshold, as also reported in the past at lower redshifts \citep{cutler_2024, hamadouche_2025}. Finally, since a possible correlation with redshift could be affecting the scatter around the median, the measurements were then split at the 50th-percentile in redshift, $ z\approx 3.5$ in both the filters. The best-fit parameters for the stellar mass-size relation in F200W and F444W are reported in Table \ref{zsplit_tab} and the models are shown in Figure \ref{fig:F200+F444_split}. In both filters, the slope across redshift bins differs by a factor of $\approx 3-4.4$, lower in the low-redshift bin, and the sizes at $M_\star= 5\times10^{10}M_\odot$ differ by $\approx 0.2$ dex. Therefore, at high redshift the variety of the population increases even though the intrinsic scatter is comparable to the low-redshift population. 
In agreement with the results in Section \ref{multiband_section}, the size gradient with wavelength is mildly negative ($\sim0.1$ dex) and the dispersion marginally ($0.01-0.02$ dex) diminishes with increasing wavelength. In both redshift bins, the best-fit results are consistent within $1\sigma$ with recent literature for smaller samples at $z=3-4$ \citep{Ito_2024, kawinwanichakij2025}. In Figure \ref{fig:z_dep_all}, we show the redshift evolution of the size at fixed $M_\star=5\times10^{10}\,M_\odot$ and we compare it with results in the literature. Since the uncertainty on the distribution of sizes is $\approx 0.2$ dex, in Figure \ref{fig:z_dep_all} one sees that the median values of effective radius in the redshift bins are roughly $1\sigma$-consistent between F200W and F444W. Our results in F200W (optical rest-frame at $z\sim3-3.5$) are broadly consistent with the extrapolation of the best-fit parameterizations traced for the bandwidth-normalized radius at $0.5\:\si{\mu m}$ in \citet{vanderwel} ($R(z)\sim5.6\times (1+z)^{-1.48}$), \citet{Straatman_2015} and \citet{Yu_2026}.  
However, our points may suggest a steeper gradient than the best-fit parameterizations in these works.
Our best-fit relations are consistent with recent determinations for similarly selected, albeit less, quiescent galaxies in \cite{Ito_2024} and \cite{kawinwanichakij2025}. Our estimate of the intrinsic scatter ($\sigma \sim 0.3$ dex) is consistent with the one in \cite{Ito_2024} and \cite{Yang}, but larger than the value in \cite{kawinwanichakij2025} based on their spectroscopic sample of 17 objects. Considering similar redshift ranges and the filter F200W, the slope of our mass-size relation is consistent with \cite{chen2026}, but our normalization is $\sim0.1$ dex lower and our intrinsic scatter $\sim0.1$ dex larger. \citealt{chen2026} also report a bending of the stellar mass-size relation at $\log(M_\star/M_\sun)=10.0$. When performing our analysis with this lower limit, the slope is decreased by $\sim 50\%$ but normalization and scatter remain consistent within $1\sigma$. Finally, we verified that the stellar mass-size relation does not change within $\sim 1\sigma$ after randomly removing 20\% of the objects in the classical UVJ-box ($U-V > 1.3$, see Fig. \ref{fig:goodSelUVJ}) -- a percentage similar to the contamination of dusty SFGs in $UVJ$-selected QGs \citep{Schreiber_2018}. Even removing the 20\% reddest sources ($V-J>1.2$) closest to the region occupied by dusty SFGs does not appreciably change the parameters of the mass-size relation. Our fiducial relation is also robust against the removal of the S/N cut described above.
\begin{table}[h]
    \centering
    \caption{Best-fit parameters}
    \begin{tabular}{c|ccc} 
     \toprule
     Filter & Slope $\alpha$  & $\log A$ & Scatter $\sigma_{\log R_e}$\\
      \midrule
     F200W($z > 3.5$) & $0.62^{+0.15}_{-0.16}$ & $-0.32^{+0.04}_{-0.04}$ & $0.31^{+0.04}_{-0.03}$ \\ [0.5ex] 
     F200W($z \leq 3.5$) & $0.14^{+0.14}_{-0.09}$ & $-0.10^{+0.04}_{-0.04}$ & $0.29^{+0.03}_{-0.03}$ \\ 
     \midrule
     [0.5ex] 
     F444W($z > 3.5$) & $0.59^{+0.14}_{-0.14}$ & $-0.39^{+0.04}_{-0.04}$& $0.27^{+0.03}_{-0.03}$ \\
     [0.5ex] 
     F444W($z \leq 3.5$) & $0.21^{+0.15}_{-0.13}$ & $-0.19^{+0.04}_{-0.04}$& $0.28^{+0.03}_{-0.03}$ \\[1ex] 
     \bottomrule
    \end{tabular}
    \tablefoot{$z$-binned best-fit parameters of the size-mass relation. Diagrams are given in Figure \ref{fig:F200+F444_split}.}
    \label{zsplit_tab}
\end{table}

\subsection{The size of the highest-redshift quiescent candidates}

In our selection, we retrieve the spectroscopically confirmed targets at $z>4.5$ in \citet{carnall2024} and \citet{degraaff_2024} (ID \#28 and 3 in Table \ref{tab:catalog}).
Our size estimates are consistent with those in \citet{carnall2024} and \citet{degraaff_2024}, while stellar masses differ up to $0.2$ dex, consistent with typical systematic uncertainties due to the different approaches to SED modeling. In this work, we also selected $3$ additional quiescent galaxy candidates at $z\sim5-6.3$, labeled by their redshift in Figure \ref{fig:F200+F444_split}. Their stellar masses from \texttt{eazy-py} range $10^{10.4-10.6}\,M_\odot$, yet above the threshold set to compute the mass-size relation ($M_\star > 10^{10.3}\,M_\odot$). Their sizes (mean and standard deviation $R_{\rm e}=0.122\pm0.002$ kpc) are also $2-8\times$ more compact than those in the literature (including those in \citealt{baker2025d}), placing them at the lower end of the mass-size relation at $z>3.5$ in the F444W filter, yet roughly within its intrinsic scatter. 

The very compact sizes might be suggestive of possible contamination of point-like sources, such as brown dwarfs \citep{Yang_2026, Kauffman_2020} or ``little red dots'' \citep{Matthee_2024}. To investigate the possible impact of point-like sources on our results at high redshift, we modeled all sources at $z>3.5$ as PSFs.
In general, using the BIC criterion, we find $BIC(\mathrm{point\:source}) - BIC(\mathrm{single\:Sersic}) < -100(10)$ in less than 5\%(10\%) of the sources, meaning that the point-source model performs better than a S\'{e}rsic profile for a small fraction of the sample. The mass–size relation without the candidate point sources does not differ systematically from our fiducial relation based on the full sample. We then turned to commonly color selections to establish the physical nature of the possible compact contaminants. Applying the cut for LRDs in \cite{Greene_2024}, $\sim 10\%$ of our $z>3.5$ satisfies the criterion, but, notably, only $\sim50\%$ of the point sources are classified as LRDs, highlighting how there are bona fide compact red quiescent candidates at high-redshift. Focusing on the three highest redshift candidates in our sample, they all meet the LRD color criteria, but only 1/3 is better modeled as a PSF than a S\'{e}rsic profile ($z_\mathrm{phot} = 6.32$). Finally, by applying the color criteria in \citet{Langeroodi_2023}, we find F150W-F200W colors $>0.25$ for all three high-redshift sources, therefore disfavoring the brown dwarfs solution. Overall, these tests further confirm the quality of the fits and the purity of the sample, ensuring minimal contamination from point-like sources.
While waiting for definitive proof of the nature of the most compact, red, and most distant galaxies, we thus caution that our claims need to be accepted with reservations. Also in this case, for full transparency the SEDs of these galaxies are released as supplementary material along with all those of the whole sample.

\subsection{The intrinsic scatter of the stellar mass-size relation}
\label{single_sersic}
 
Our large statistical sample of robust size measurements allows for determining the intrinsic scatter of the stellar mass--size relation, which is rather constant ($\sigma\sim0.3$ dex) in the two redshift bins under consideration and in both filters within the uncertainty (Figure \ref{fig:F200+F444_split} and Table \ref{zsplit_tab}).
At high redshift the size change with mass is steeper than at lower redshift (Table \ref{zsplit_tab}), testifying to a high accretion efficiency at earlier epochs. Moreover, galaxies at higher redshift show smaller sizes at a given mass by $\sim 0.2$ dex in F444W and $\sim 0.22$ dex in F200W with respect to lower redshift. The universe was denser the earlier the epoch under consideration and this seems to be inherited by the galaxies' evolution rate. The galaxies who synchronously formed at earlier epochs experienced a steeper size change as a function of mass. In two given volume shells $\partial\mathcal{V}(\left[z_0, z_1\right])$, $\partial\mathcal{V}([z^\prime_0, z^\prime_1])$ with $z_i < z_i^\prime$, $i = 0, 1$, if $\Delta \log(R)/\Delta \log(M_\star/M_\odot)$ is lower in $\partial\mathcal{V}(\left[z_0, z_1\right])$ it means that there is a homogenization of size in the galactic population within a given mass range as time goes on. This is also in agreement with \citet{hamadouche_2025}.

We note that the intrinsic scatter is partially driven by the presence of some particularly large ($R_e \sim 1-3.5\:\mathrm{kpc}$) galaxies consistently in all filters, and especially at  $z\leq 3.5$  (Figure \ref{fig:F200+F444_split}), affecting both slope and intercept of the best-fit relation. These sources show good residuals with simple single-Sérsic profiles, small error bars, and their photometric redshift is well constrained. Recent work reported the existence of a few quiescent galaxies at $z\sim3-5$ and relatively low-mass ($M_\star = 10^{9.5}- 10^{10.3}M_\odot$, \citealt{sato_2024, baker2025d}), similar to the occurrences in our sample close to $M_\star \gtrsim 10^{10.3}M_\odot$.
We further investigated the possible spurious origin of such large radii in our galaxies. Close pairs and mergers might artificially inflate galaxy sizes. As mentioned in Section \ref{pysersic}, 11 close companions or galaxies with extended, unresolved (see, e.g. \ref{fig:F444W_resids}) features were simultaneously re-modeled with manual prior. Nine out of eleven sources could not be properly deblended and were excluded, while in 2/11 galaxies this new fit resulted in a decrease of $14$\% and $50$\% of the effective radius, which does not explain the still large sizes observed in galaxies above the stellar mass-size relations in Figure \ref{fig:F200+F444_split}. 
We return on the implementation of more complex two-component models in the next Section. However, barring unusually high incidence of such contamination of unaccounted morphological components, the presence of large QGs at $z>3$ is an intrinsic property. 

\begin{figure*}
     \centering
     \includegraphics[width=1\linewidth]{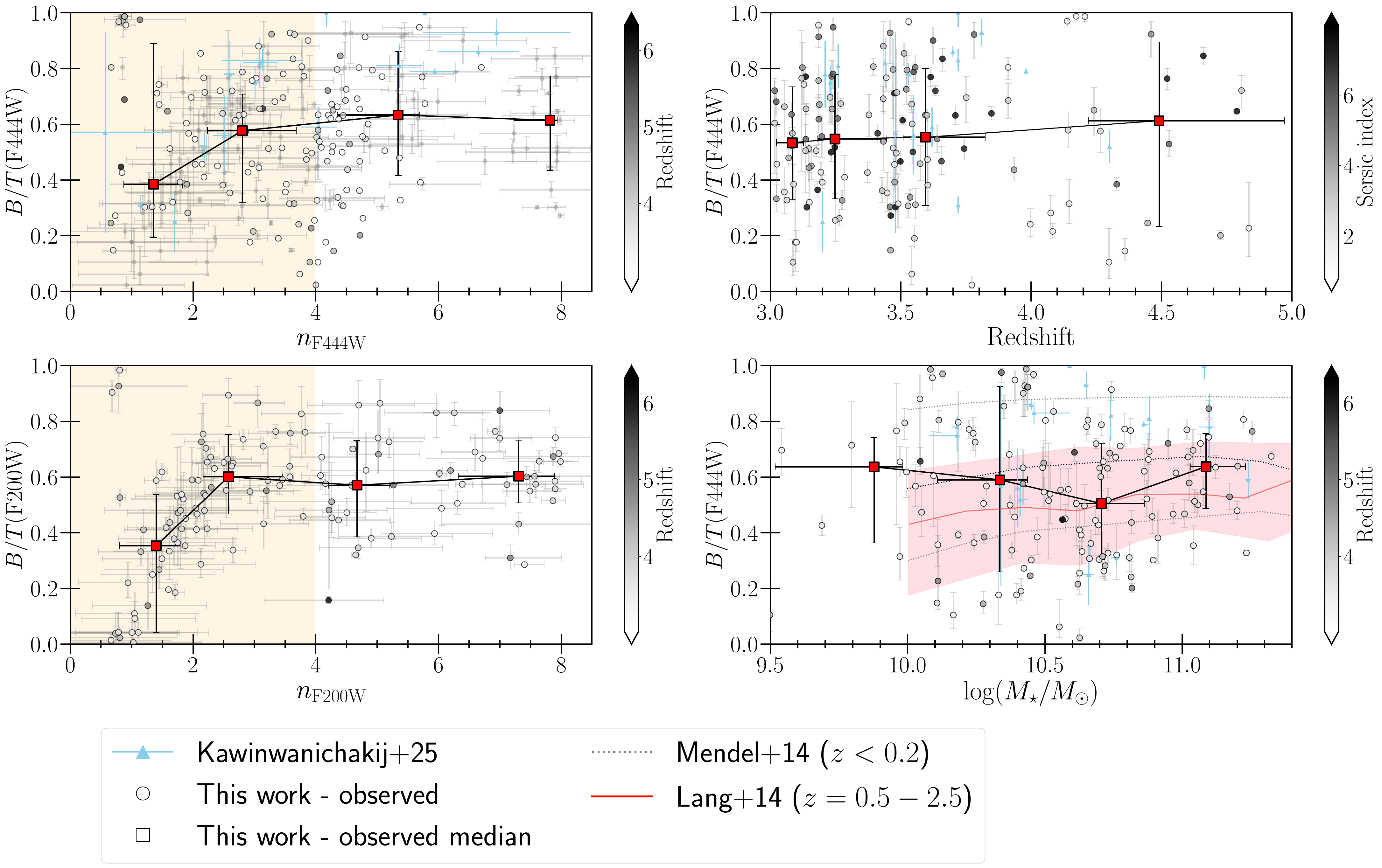}
     \caption{Bulge-to-total ratio as a function of Sérsic indices from a simple S\'ersic profile (in the F444W, top-left, and F200W bands, bottom-left), redshift (top-right), and stellar mass (bottom-right). Gray circles, color coded as indicated in each panel, represent the measurements in this work. The light orange region indicates $n\leq4$. Light blue triangles mark the measurements in \citet{kawinwanichakij2025}. Our median values are shown as red squares. The red solid line and shaded area in the bottom right panel are from \citet{Lang_2014}, while the gray dotted lines are from \citet{Mendel_2014}.}
     \label{fig:BT_ratio}
 \end{figure*}

\subsection{Beyond simple S\'{e}rsic profiles: a bulge-disk decomposition}
\label{bulgedisk}
So far, we analyzed results based on the assumption that a simple S\'{e}rsic profile is sufficient to accurately describe the surface brightness of QGs at high redshift. This allowed us to compare our results with literature at lower redshifts. However, recent studies highlight potential discrepancies in fit quality and outcome when the bulge and disk components are treated as distinct contributors to the surface brightness of QGs. As reported in \citet[][]{kawinwanichakij2025}, the bulge-to-total ratio ($B/T$) does not attain unity for all UVJ-selected quiescent galaxies. The presence of large disks could also explain some of the large radii mentioned above.\\

To address this, we compared the quality of simple Sérsic profile models against that of a bulge-disk decomposition. We modeled the surface brightness of our sample with two S\'{e}rsic profiles with \texttt{pysersic}, one fixed to an exponential disk with $n=1$ (\texttt{sersic\_exp} in \texttt{pysersic} nomenclature). We allowed the code to automatically determine the priors, masking contaminants with \texttt{SEP}, and we initialized the disk effective radius to be $1.5\times$ larger than for the other S\'{e}rsic component.
For the fiducial sample of 129 galaxies
with well determined sizes in all filters from single S\'ersic models, and taking F444W as a reference band, the two-component fit returns a median bulge to disk size ratio of
$R_\mathrm{Bulge,\, F444W}/R_\mathrm{Disk,\,F444W} \coloneqq \mathcal{R}_\mathrm{F444W} = 0.46^{+2.77}_{-0.34}$ ($0.19^{ +0.31} _ {- 0.07}$ for galaxies with bulge size smaller than disk size), where the uncertainties reflect the 16-84th percentile range.
The average bulge-to-total flux ratio is $B/T_\mathrm{F444w} = 0.56 \pm 0.24$. Both estimates are consistent with the results reported by \citealt{kawinwanichakij2025} within the uncertainties.

We applied the BIC criterion to establish whether a bulge-disk decomposition constitutes an improvement over simple Sérsic profile models. We used the BIC criterion to compare the performance of the two models. For completeness to the classical convention $\Delta BIC < -10 $, we also report the result for a more stringent criteria $\Delta BIC < -100 $ and define $BIC(\mathrm{bulge+disk}) - BIC(\mathrm{single\:sersic}) = \Delta BIC < -100(10)$ to stress the difference between the efficiencies of the two profile types. Since we are dealing with faint, red sources, we require the two models under comparison to yield significantly different results to make sure that critical features are adequately modeled, if present. For $\approx 22\%(55\%)$ of the sources in our sample, the bulge-disk decomposition performs better than a single Sérsic profile. Even in the case that a higher percentage were to be found, however, the \texttt{sersic\_exp} assumes in the prior that $n_\mathrm{disk} = 1$ while $n_\mathrm{bulge}$ is left free to vary. There are sources for which the Sérsic index in Table \ref{tab:catalog} is $n<4$ and this suggests that the radius of the disk component is likely overestimated. To characterize this trend, we studied the dependence of the bulge fraction $B/T$ as a function of: Sérsic index $n$, redshift, and stellar mass $\log (M_\star/M_\odot)$ with resulting intrinsic scatters 
 $\sigma_\mathrm{int, F444W} = 0.21, 0.23, 0.22$ dex 
 respectively (see \href{https://zenodo.org/records/18669101?preview=1&token=eyJhbGciOiJIUzUxMiJ9.eyJpZCI6ImMzYmNlODUzLWQwYTktNDdiYi05ZDQyLTBiMWU1YzI5YjJlOSIsImRhdGEiOnt9LCJyYW5kb20iOiJkNDM4YjkzNzY4ZWZiODM3NDE4MmU3OGI2MDNmMjMxOSJ9.IVMoA4i5csQY9J1VSCZTYVXtkbNN1kredIkizICWpxqamlRAXkk4Y69LNf60mjjIHZ8KeEgEbEyGbyC0lFAXcA}{Zenodo}).
We notice that 35\% of the sources shows a disk size smaller than the bulge size, which would explain the large uncertainty in $\mathcal{R_\mathrm{F444W}}$. 
In this case, the bulge-disk decomposition is not supported by strong evidence from the data.
Results can be seen in Figure \ref{fig:BT_ratio}. The median B/T ratios as a function of \mstar\ are plotted together  with the estimates from \citealt{kawinwanichakij2025} and literature trend at lower redshifts \citealt{Lang_2014, Mendel_2014}. 
One can see a mild increasing trend for $B/T$ driven by low Sérsic index, but for stellar mass and redshift no correlation can be detected.

We thus conclude that for the majority of our sample, a single S\'{e}rsic profile is sufficient to model the surface brightness, lending credence to the analysis reported in the previous section.

 \section{Predicting the size of distant QGs}
 \label{result_analysis}
In the previous sections, we addressed the classical stellar mass--size relation for QGs at $z>3$, the highest redshift range for which we could assemble a large statistical sample. In this section, we tackle a different question and, with the aid of Bayesian statistics, identify the observables or physical quantities that best predict the sizes of distant quenched galaxies, thereby disentangling their contribution to large intrinsic scatter on the mass--size relation.

 \begin{figure*}
     \centering
     \begin{subfigure}[b]{0.48\textwidth}
         \centering
         \includegraphics[width=\textwidth]{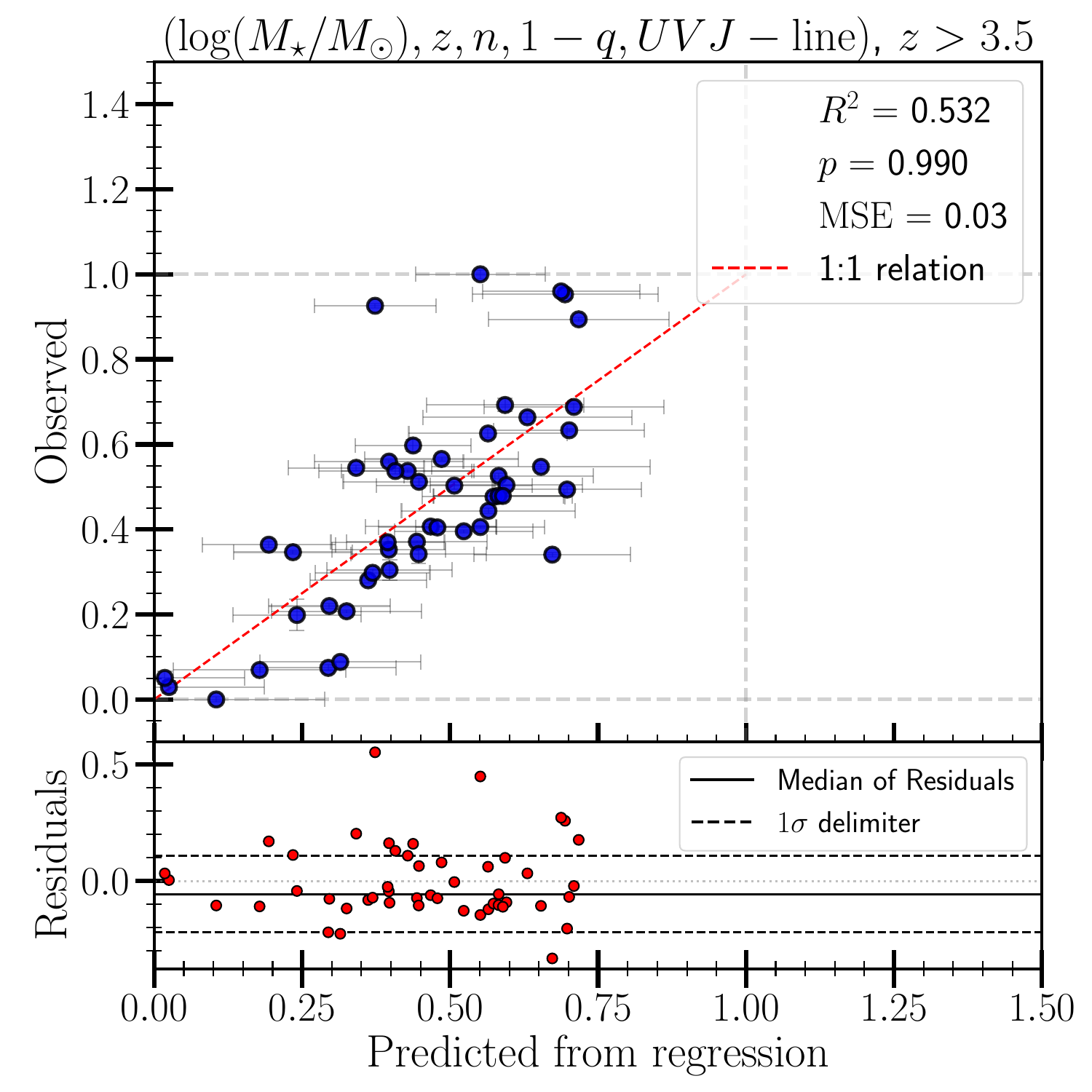}
         \caption{}
         \label{fig:finalforwardupper_2}
     \end{subfigure}
     \hfill
     \begin{subfigure}[b]{0.48\textwidth}
         \centering
         \includegraphics[width=\textwidth]{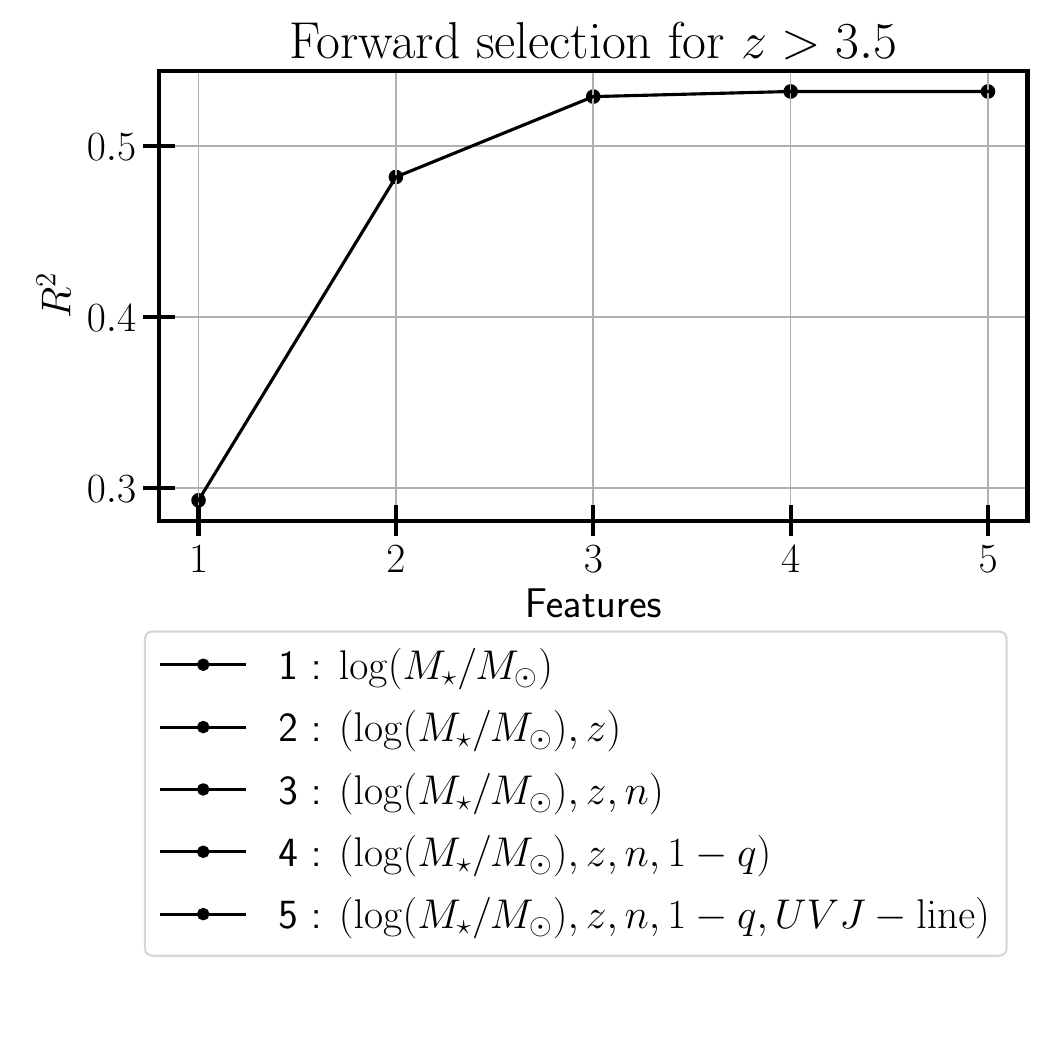}
         \caption{}
         \label{fig:finalforwardupper_3}
     \end{subfigure}
    \vspace{\baselineskip}     
    \begin{subfigure}[b]{0.48\textwidth}
         \centering
         \includegraphics[width=\textwidth]{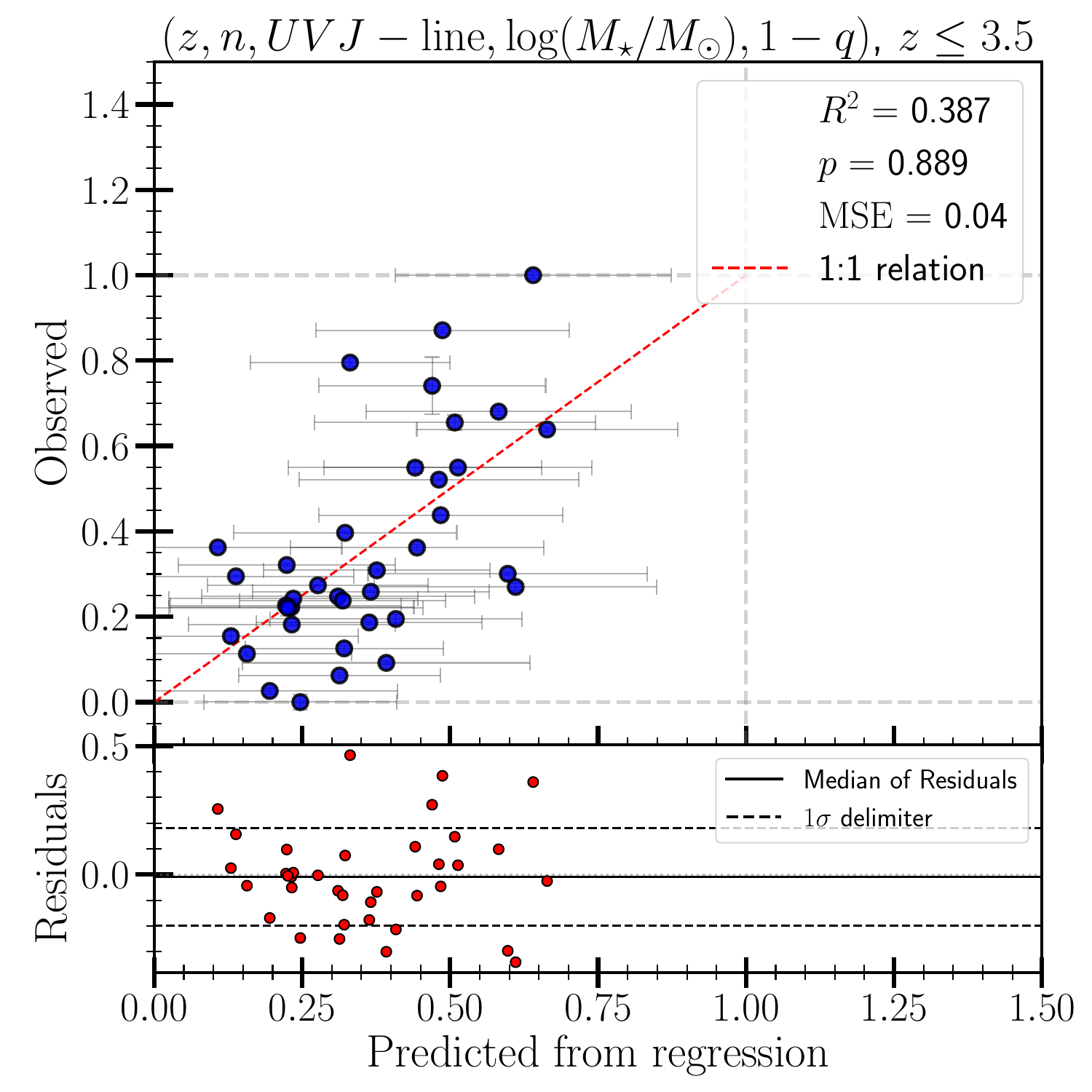}
         \caption{}
         \label{fig:finalforwardbottom_1}
     \end{subfigure}
     \hfill
     \begin{subfigure}[b]{0.48\textwidth}
         \centering
         \includegraphics[width=\textwidth]{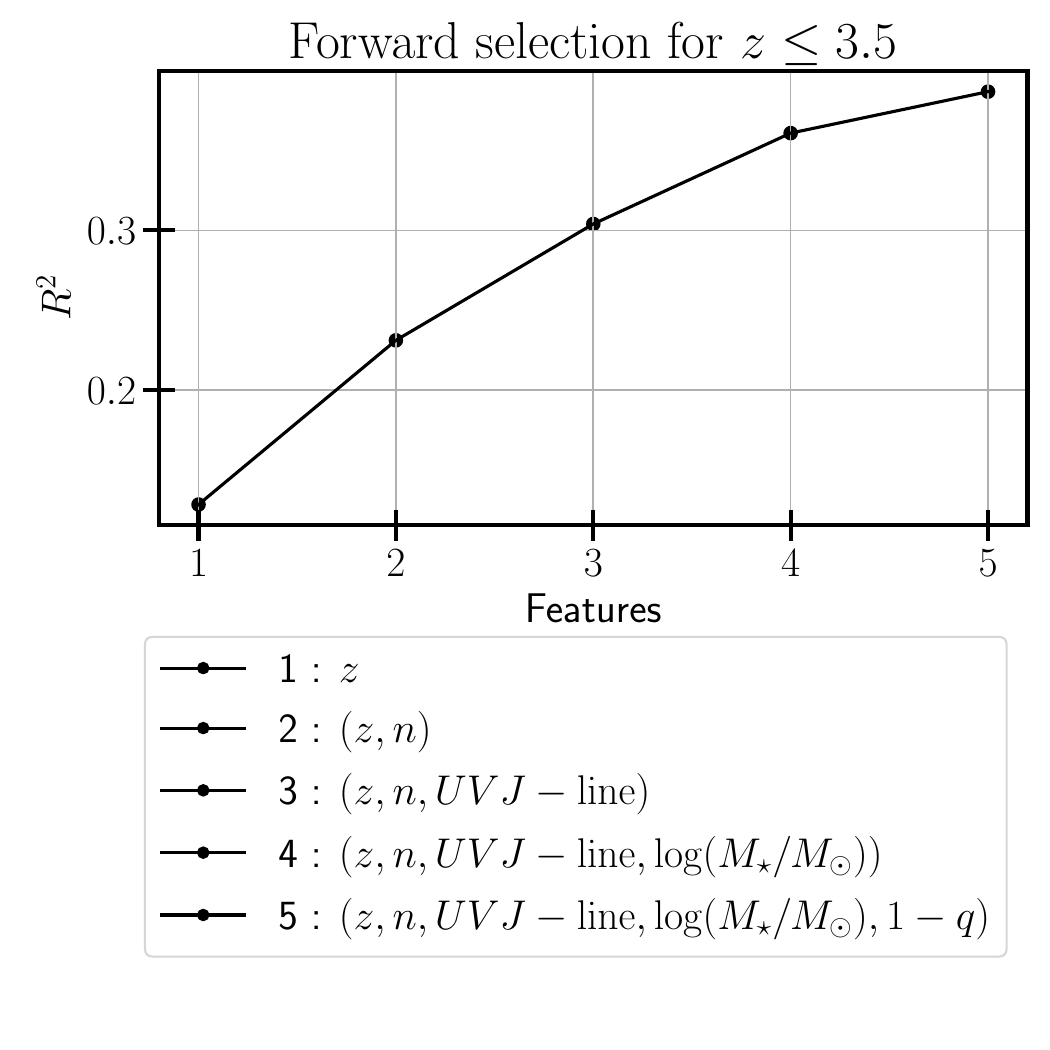}
         \caption{}
         \label{fig:finalforwardupper_4}
     \end{subfigure}
        \caption{Panel (a): One-to-one plot between predicted values from regression and observed values of the effective radius obtained using a linear combination of parameters. The sources come from the high-redshift bin. The blue points show the points computed with the full linear relation whose variables are noted at the top of the plot. The red points in the lower part of the figure show the residuals between the best-fit and observed quantities. Panel (b): Evolution of the $R^2$ value during the forward selection. The values are from sources in the high-redshift bin. The variables are sorted in descending order of importance from the left. The x-axis reports the number of variables used for the fit (or the step of the procedure). Panel (c): Same as in panel (a), but for the low-redshift bin. Panel (d): Same as in panel (b), but for the low-redshift bin. In the upper right corner of panels (a) and (c), we give $p$ and MSE.} 
        \label{fig:FinalForwards_2}
\end{figure*}

\subsection{Method}
\label{param_analysis}
Here we adopt an approach analogous to \citealt{runnholm2020}. Given a pool of $N$ parameters, we model their relation with the logarithm of the effective radius using a series of linear models with progressively increasing or decreasing number of variables. We model the dependence of the effective radius on the parameters as:
 \begin{equation}
 \label{fit_eq}
     \log \left[R_e(\mathbf{X})\right] = \mathbf{\theta}^\mathrm{T}\mathbf{X} + b
 \end{equation}
where $\mathbf{\theta}$ is the vector of coefficients, $b$ is the intercept, $\{\mathbf{X}_i\} = \{X_{1, i}, ..., X_{k, i}\}$ are the vectors of the $k$ independent parameters (features) for each $i$th object. To assess the goodness of fit of each model and determine whether the addition or removal of a specific parameter improves the description of the sizes, we rely on the use of the weighted $R^2$ metric, defined as:
 \begin{align}
     R^2 &= 1 - \frac{\sum_i(y_i - \hat{y}_i)^2}{\sum_i(y_i - \mathbb{E}[\mathbf{Y}])^2}
 \end{align}
 where $\mathbf{Y}$ are the sample points, $\hat{\mathbf{Y}}$ the best-fit estimate from the model, and $\mathbb{E}[\mathbf{Y}]$ the sample mean.
 We then applied the forward (backward) selection of variables based on the maximum (least decreased) $R^2$ value\footnote{We note that \texttt{scikit-learn} allows the use of $R^2$ with sample weights, which should be used together with the total variance of the data. Otherwise, $R^2$ can become negative for highly scattered data.} associated with each parameter combination. In practice, the relation in Eq. \ref{fit_eq} is fitted in the forward selection for all variables taken in groups of $k$ ($\mathrm{dim}\left[\Theta\right] = k$), where $k$ is the step of the process. For $k = 1$ (single variable fitting), we select the model with the variable with the highest $R^2$ value. The corresponding parameter is then kept in the model while all remaining variables are tested adding one at a time for $k = 2$, and so on. For $N$ total parameters, there are $N - k + 1$ possible groups at each step. For the backward selection we start with a model that includes all parameters simultaneously and progressively remove one variable at a time, each time selecting the one whose removal yields the smallest decrease in the $R^2$ value. 

\subsection{Choice of variables and normalization}
\label{parameters_analysis_F444}
The design of the parameter space $\Theta$ requires to choose quantities relevant to the evolution of the galaxy size. We chose the logarithmic mass $\log (M_\star/M_\odot)$ along with Sérsic index $n$ and ellipticity $1-q$ as direct tracers of the morphology and the logarithmic redshift $\log(1+z)$ as describing the cosmological stage of the universe hosting our galaxies. We then include the linear combination of the \textit{UVJ} colors mapping the location of our targets in this parameter space. 
We follow \citealt{williams2009} and \citealt{Ito2025} and define the following coordinate transformation:
 \begin{align}
     f(U, V, J) = 0.75(V-J) + 0.66(U-V)
 \end{align}
 which tracks the position along the diagonal line of the selection box for quiescent galaxies (Figure \ref{fig:goodSelUVJ}), broadly mapping the aging of the stellar populations since quenching (with dependencies on other factors, \citealt{Cheng_2024}). We opt not to include physical quantities such as the star formation rate (SFR) or sSFR because their absolute values for truly quiescent galaxies are not reliable due to the uncertainties of photometric SED fitting, although sufficient to establish their nature.  It is customary to write variables in a Gaussian form, but the distributions of our selected parameters show that they are not. We thus opt to normalize all the random variables $X$ to values between 0 and 1: 
 \begin{align}
 \label{normalization}
     &\phi(X) = \frac{X - \mathrm{min}\{X\}}{\mathrm{max}\{X\} - \mathrm{min}\{X\}}
 \end{align}
 which changes their uncertainties by the same simple multiplicative factor. We also tested non-normalized variables, repeating the entire process, but no discrepancy was found in terms of predictive power of the variables.
 \subsection{Results}
To interpret the results of this section in the context of the large intrinsic scatter on the mass-size relation (Figures \ref{fig:F200+F444_split} and \ref{fig:z_dep_all}), we opt for splitting the sample according to the same redshift bins ($3<z<3.5$, $z\geq3.5$) and applying a stellar mass threshold. For $z > 3.5$ ($z \leq 3.5$), having mass cut $\log(M_\star/M_\odot) > 10.30$ (10.5)\footnote{Due to difficulties in estimating the posterior, caused by highly scattered data and many variables, we chose the minimum mass cut that was high enough to obtain a result.} we modeled the properties of 49 (37) objects, respectively. Figure \ref{fig:FinalForwards_2} shows the results of our parameter analysis for sizes measured in the F444W filter as a reference, along with the mean squared error (MSE) and the $p$-value ($p$) (see also Fig. \ref{fig:singleVar_fits} for $k=1$). 

In the $z>3.5$ bin, the stellar mass is the parameter that affects the most our fitting, therefore representing the best predictor of the size evolution, followed by the redshift (possibly due to the large redshift range encompassed by this bin), S\'ersic index, ellipticity, and UVJ colors. The backward variable selection returns similar results. The combination of all parameters returns a total $R^2 \sim 0.5$, whereas the large $p$-value is driven by the large intrinsic scatter. We note that the improvement on $R^2$ after including \mstar\ and $z$ is minimal: adding extra features among the ones considered here does not explain a significant fraction of the scatter.    
In the $z \leq 3.5$ bin, the situation is not clear. The scatter is such that, even combining all the parameters, we obtain an even lower total $R^2 \sim 0.4$, with minimal improvement when adding one parameter at the time. The leading variable is redshift, while the stellar mass comes after the S\'ersic index and the $UVJ$ colors. Possibly due to the reduced mass range and small sample statistics, our experiment returns inconclusive results in this case.

As a counter-check, we performed a partial correlation analysis \citep{Bluck_2020, Baker_2022} between residuals of effective radius and parameters at each step. We find that the redshift and the Sérsic index have the most significant partial correlation coefficients\footnote{Computed with \texttt{pingouin} library in Python, using the function \texttt{partial\_corr}} of $0.34$ ($p = 0.001$) and $-0.18$ ($p=0.077$), respectively (Figure \ref{fig:FinalForwards_app} in Appendix for the residuals and \ref{fig:paramtoparam} for cross-dependencies).
In general, the size residuals $\Delta R(\theta)$ do not show strong, clear trends as a function of each of the other individual parameters, supporting the results of the variable selection described above. This implies that the current scatter cannot be explained by the set of parameters considered here.


\subsection{Looking for possible environmental effects}
Finally, we attempted to extend the pool of possible variables affecting quiescent galaxy sizes at $z>3$ to include their environment \citep{kawinwanichakij2025}. Estimating the environment, especially at high redshift, is notoriously subject to several assumptions implicit in each observation techniques (see \citealt{kakimoto_2024, kakimoto_2026a, kakimoto_2026b, mcconachie_2025, binh_2025} for recent attempts for distant QGs).
For simplicity, here we focus on the subsample of sources in the CEERS field, hosting one of the largest and strongest overdensities of galaxies at $z=3.44$, and that is single-handedly responsible for elevated number densities of quiescent galaxies \citep{carnall_2023, valentino2023ApJ, jin_2024, Sillassen_2025, Ito_2025_merger}. 
We recomputed the overdensity significance of galaxies in CEERS as detailed in \cite{baker2025d, Ito_2026_lowmass}. Briefly, we considered sources in the \texttt{eazy-py} catalogs (Section \ref{data}) and applied a stellar mass cut at $\log (M_\star/M_\odot) > 8$ and a redshift quality cut $(z_\mathrm{phot, 84} - z_\mathrm{phot, 16}) / z_\mathrm{phot, 50} < 1$, where $z_\mathrm{phot, q}$ is the $q$th percentile of the probability distribution of the photometric redshift. We considered a redshift slice at $z=3.4$ with $\delta z = 0.1$ width and smoothed over a radius of $1$ Mpc. We derived an average galaxy density map in aperture of 500 pkpc and computed the overdensity around each candidate QG as $(N-\langle N \rangle)/\langle N \rangle$, where $N$ is the number of galaxies within the aperture and $\langle N \rangle$ the average. We then repeated the analysis in Section \ref{parameters_analysis_F444} and found no impact of the overdensity parameter on the galaxy sizes, with $R^2 \sim 10^{-2}$ when directly contrasting $R_{\rm e}$ and the overdensity parameter. This is consistent with the analysis of the mass--size relation for the spectroscopically confirmed QG in the protocluster in \cite{Sillassen_2025}. Therefore, using the extreme environment detected in this field as a stress test on the environmental effects on our sample, we do not find a significant impact on the size of quiescent galaxies at $z\sim3.5$. A full-fledged analysis on the overall sample, or at least the portion benefiting from a systematic spectroscopic follow-up on large scales and homogeneous data for the density reconstruction as that available for the Cosmic Vine studied here, will test our findings.

\section{Discussion and conclusions}
\label{discussion_summary}

In this work, we present a new observational benchmark for morphological studies of quiescent galaxies at $z>3$, exploiting the sensitivity and spatial resolution at rest-frame optical wavelengths enabled by JWST/NIRCam. We model the multi-wavelength emission of $\sim130$ color-selected quiescent galaxies distributed over $825$~$\mathrm{arcmin^2}$ across several well-studied extragalactic fields. Using classical Sérsic profile fitting and bulge–disk decompositions, we derive morphological parameters and trace the stellar mass--size relation at $z>3$, and, as a key result of this study, robustly ($\lesssim 0.1$ dex uncertainty) constrain its large intrinsic scatter of $\sim0.3$ dex. We further investigate the presence of wavelength-dependent gradients in the derived morphological properties. In the following, we summarize and discuss our main results.

\begin{itemize}
    \item we prove that, on a population level, the size mildly decreases with increasing wavelength, whereas for the Sérsic index we do not see any relevant dependence on the filter used. The size gradient with stellar mass makes more massive sources appear smaller at longer wavelength while for the Sérsic index gradient we see no indication of any dependence on the stellar mass. These effects may be explained by considering that the older, redder population still slightly dominates the emission about the center of the galaxy and reduces the apparent size at longer wavelength, suggesting inside-out quenching as a dominant mechanism on a population level; in contrast, as seen from the multi-band analysis the "bulginess" appears approximately constant across filters and to be roughly the same at all masses. This, however, is not smoothly linked to the findings at $z \leq 2.5$ \citep[see e.g.][]{martorano2025}.
    
    \item At $z \ge 3$, the normalization of the mass-size relation of quiescent galaxies keeps decreasing: the size decreases with increasing redshift at fixed stellar mass, particularly highlighting a steeper size change with mass at higher redshift. We also finally estimate a large intrinsic scatter of $\approx 0.3$ dex at all redshifts $z>3$ and wavelengths. The quiescent-galaxy SMF at $z>3$ peaks at $\approx 10^{10.5}M_\odot$ and declines toward both lower and higher masses, but with a much shallower fall-off on the low-mass side than on the high-mass side, resulting in an asymmetric distribution in stellar mass \citep{baker2025b}. This asymmetry may enhance the rate of merging events, especially of satellite galaxies, and spur further any quenching evolution already ongoing as well as mass--size evolution \citep{puskas_2025, Nipoti2025}. The progressive flattening of the mass-size relation (see \citealt{Ito_2024}) can have different potential causes. Since in our lower redshift bin $z=3-3.5$ we have a $0.14$ dex lower mean mass and we know that galaxies in transition from a star-forming state to quiescence have larger median sizes, provided this effect holds at $z>2.5$ it can: 1) cause the flattening of the mass-size relationship (Sec. 6 in \citealt{Ito_2024}) and 2) increase the scatter around the best-fit line (see Figure \ref{fig:F200+F444_split}). Indeed, our UVJ selection allows for the inclusion of diskier transitioning galaxies. If these sources are found in the green valley and are larger than bona fide quiescent galaxies at the same redshift and $n\gtrsim4$, the flattening of the relation occurs, witnessing an effect due to progenitor bias. It is worth noticing that, contrary to our results, \citealt{vanderwel} does not find a relevant dependence on redshift of the slope in the mass--size relation ($z < 3$): this discrepancy may be due to differences in the selection of sources and possible contamination. Since F200W is closely tracing the rest-frame optical, the change in slope across redshift remains even for sizes computed at $\lambda = 0.5\si{\mu m}$.
    
    \item The bulge-disk decomposition clearly indicates that the bulge is responsible for more than half of the light emitted by our galaxies and that for more than a third of the sample no disk contributes to their morphological characterization. However, we find that the bulge-disk decomposition does not significantly improve the residuals when compared to simple Sérsic profile for $\approx 80\%$ of our sample, thus making the single component a good compromise to fit our QGs. The intrinsic scatter of $\sim 0.2$ dex for B/T in all functional relations does not secure the existence of a trend on a population level.

    \item  We conclude that, by leveraging a $5\times$ larger sample than previously used for similar studies, after a significance analysis performed with Bayesian methods none of the classical morphological parameters in a Sérsic distribution nor mass nor redshift can effectively explain the aforementioned intrinsic scatter. No relevant cross-correlations are found for the considered parameters. Eventually, it is evident that other physical parameters need to be taken into account to thoroughly explain the occurrence of the intrinsic scatter and constrain the size evolution of our systems.
\end{itemize}

In particular, to improve future studies of the same kind, we regard as necessary 1) the use of larger samples of galaxies to improve the statistical significance and 2) the inclusion, in the search for the best predictor, of quantities more intrinsically related to the physics of QGs, such as dispersion velocity and SFR. These, in particular, call for extensive spectroscopic studies at higher redshift.

\section{Data availability}
All plots and models obtained in this paper, complemented by additional mathematical equations and images (like neglected sources), are available at the link \href{https://zenodo.org/records/18669101?preview=1&token=eyJhbGciOiJIUzUxMiJ9.eyJpZCI6ImMzYmNlODUzLWQwYTktNDdiYi05ZDQyLTBiMWU1YzI5YjJlOSIsImRhdGEiOnt9LCJyYW5kb20iOiJkNDM4YjkzNzY4ZWZiODM3NDE4MmU3OGI2MDNmMjMxOSJ9.IVMoA4i5csQY9J1VSCZTYVXtkbNN1kredIkizICWpxqamlRAXkk4Y69LNf60mjjIHZ8KeEgEbEyGbyC0lFAXcA}{Zenodo repository}. The Github page with the pipeline used to perform the serialized morphological analysis is \url{https://github.com/gianlucaS24/Master_Paper.git}. In the GitHub page a brief description of the names of files and folders is given.

\begin{acknowledgements}
GS acknowledges the guidance provided by Georgios Magdis in finding a Master’s project that aligned with the student’s interests and proved to be an invaluable experience for academic, professional, and personal growth. GS also acknowledges the support provided by all the individuals at DTU who have contributed to the publication of this paper through their feedback and advice. We acknowledge Minju Lee for her valuable guidance on the use of the HPC at DTU.
FV, KI, AP, and PZ acknowledge support from the Independent Research Fund Denmark (DFF) under grant 3120-00043B. WMB gratefully acknowledges support from DARK via the DARK fellowship. This work was supported by a research grant (VIL54489) from VILLUM FONDEN.
This work is based in part on observations made with the NASA/ESA/CSA James Webb Space Telescope. The data were obtained from the Mikulski Archive for Space Telescopes at the Space Telescope Science Institute, which is operated by the Association of Universities for Research in Astronomy, Inc., under NASA contract NAS 5-03127 for JWST. Some of the data products presented herein were retrieved from the Dawn JWST Archive (DJA). DJA is an initiative of the Cosmic Dawn Center, which is funded by the Danish National Research Foundation under grant DNRF140. 
\end{acknowledgements}

%
\bibliographystyle{aa}
\bibliography{bib}

\begin{appendix}



\onecolumn

\onecolumn
\section{Additional figures}

Here we show some additional figures that show alternative proceedings than those presented in the paper. We also show some of the residuals resulted from the \texttt{pysersic} fit.
\begin{figure}[h]
\centering
\includegraphics[width=\textwidth]{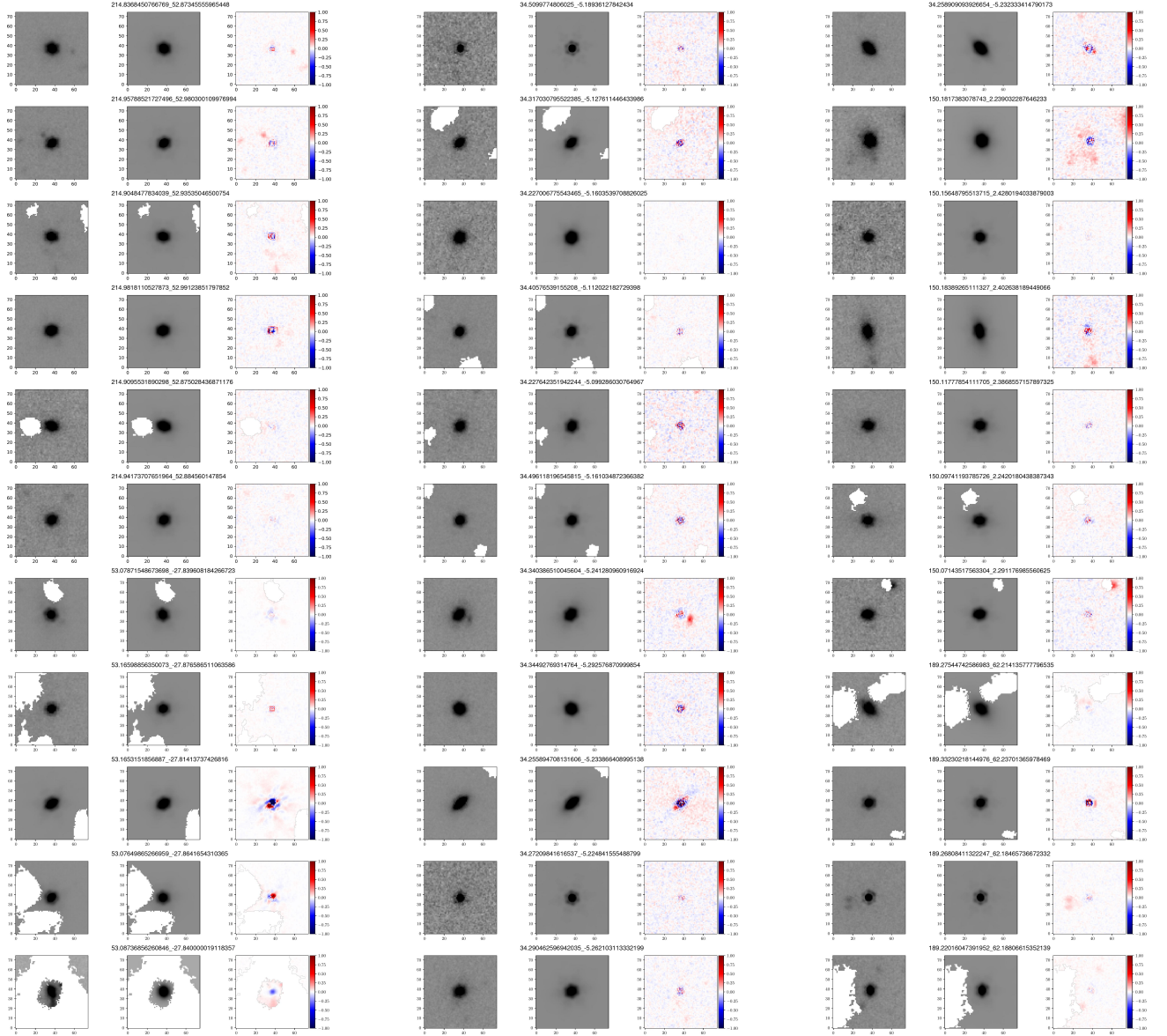}
\caption{Some residual plots in F444W where masking and single Sérsic profile is used.}
\label{fig:F444W_resids}
\end{figure}
\begin{figure}[h]
     \centering
     \begin{subfigure}[b]{0.6\linewidth}
         \centering
         \includegraphics[width=\textwidth]{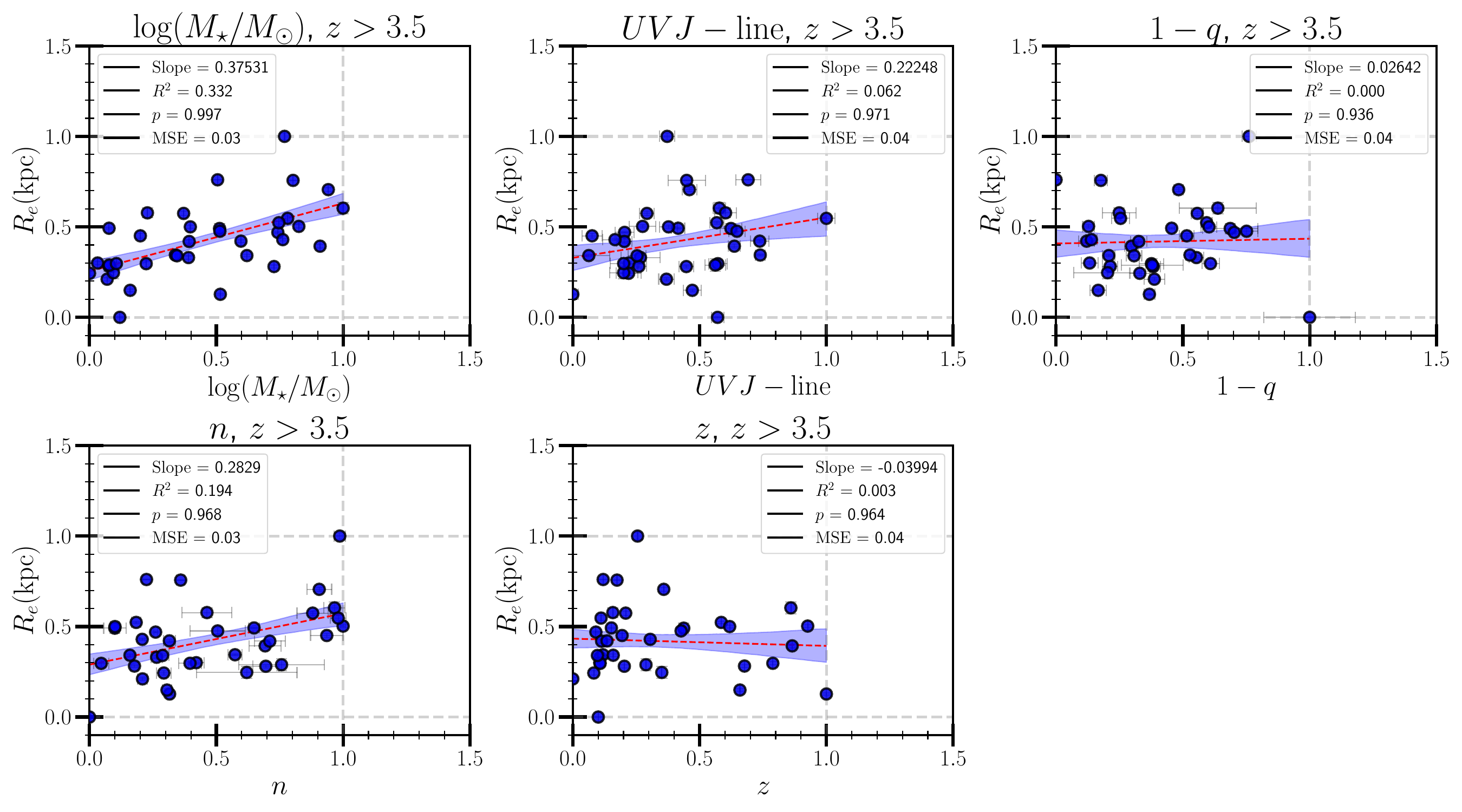}
         \caption{Fit in the high-redshift bin}
         \label{fig:features_comp_single}
     \end{subfigure}
     \begin{subfigure}[b]{0.6\linewidth}
         \centering
         \includegraphics[width=\textwidth]{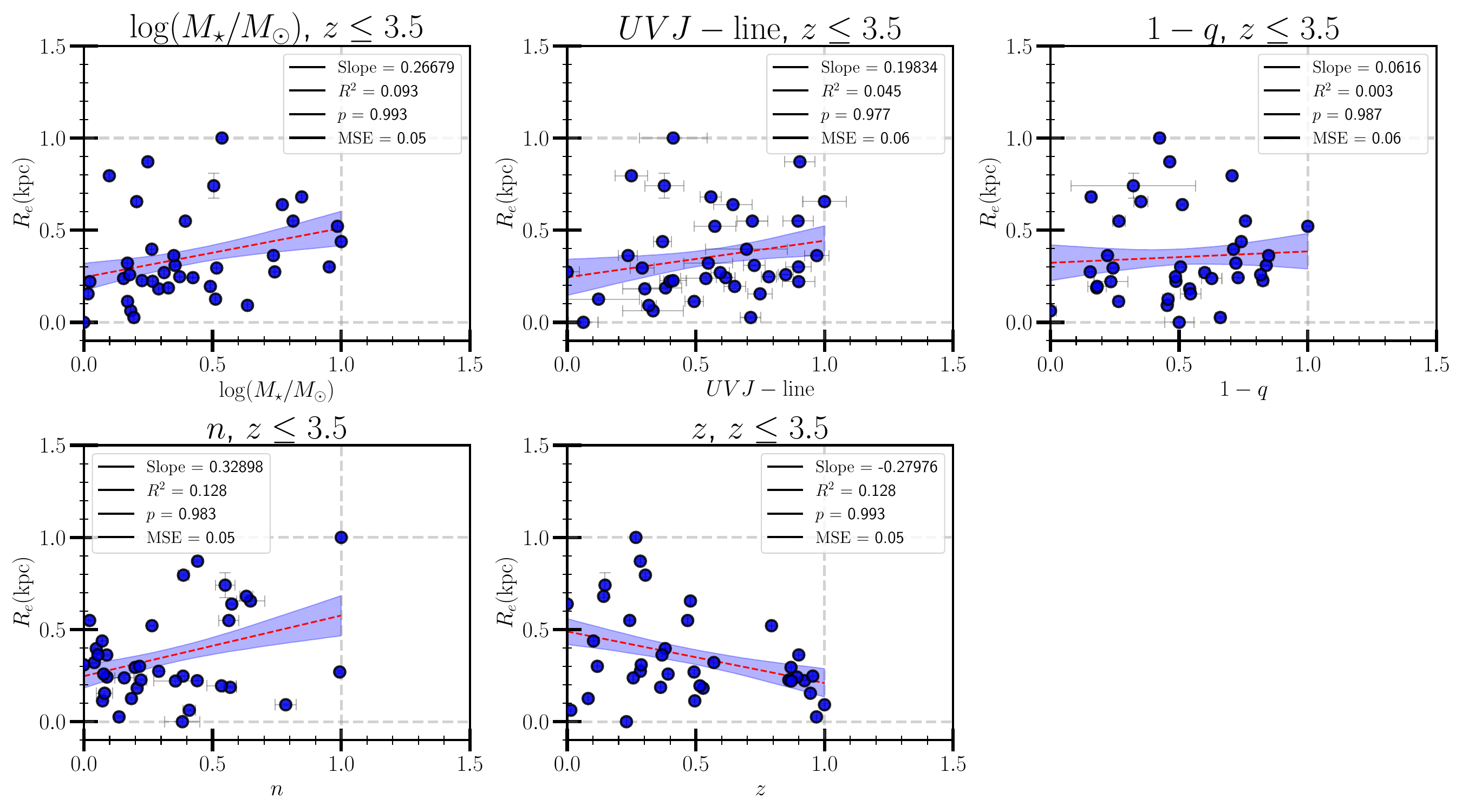}
         \caption{Fit in the low-redshift bin \textbf{with} 20\% contamination fraction.}
         \label{fig:features_compNoCuts_single_bot}
     \end{subfigure}
        \caption{Results in F444W. First step for the determination of the best predictor. The blue filled points represent the data in our sample, with the independent variable and the redshift region indicated above the plot. Red dashed lines represent the best fit to the data. The blue region is the usual $1\sigma$ region containing 68\% of the fitting line coming from the posterior estimate by \texttt{emcee}. Panel (a): Shows the the first step corresponding to Fig. \ref{fig:finalforwardupper_2}. Panel (b) Shows the fit when 20\% of the sample in the low-redshift bin is removed randomly to account for potential contamination.}
        \label{fig:singleVar_fits}
\end{figure}
\begin{figure}[h]
    \centering
    \begin{subfigure}[b]{0.48\textwidth}
        \centering
        \includegraphics[width=\textwidth]{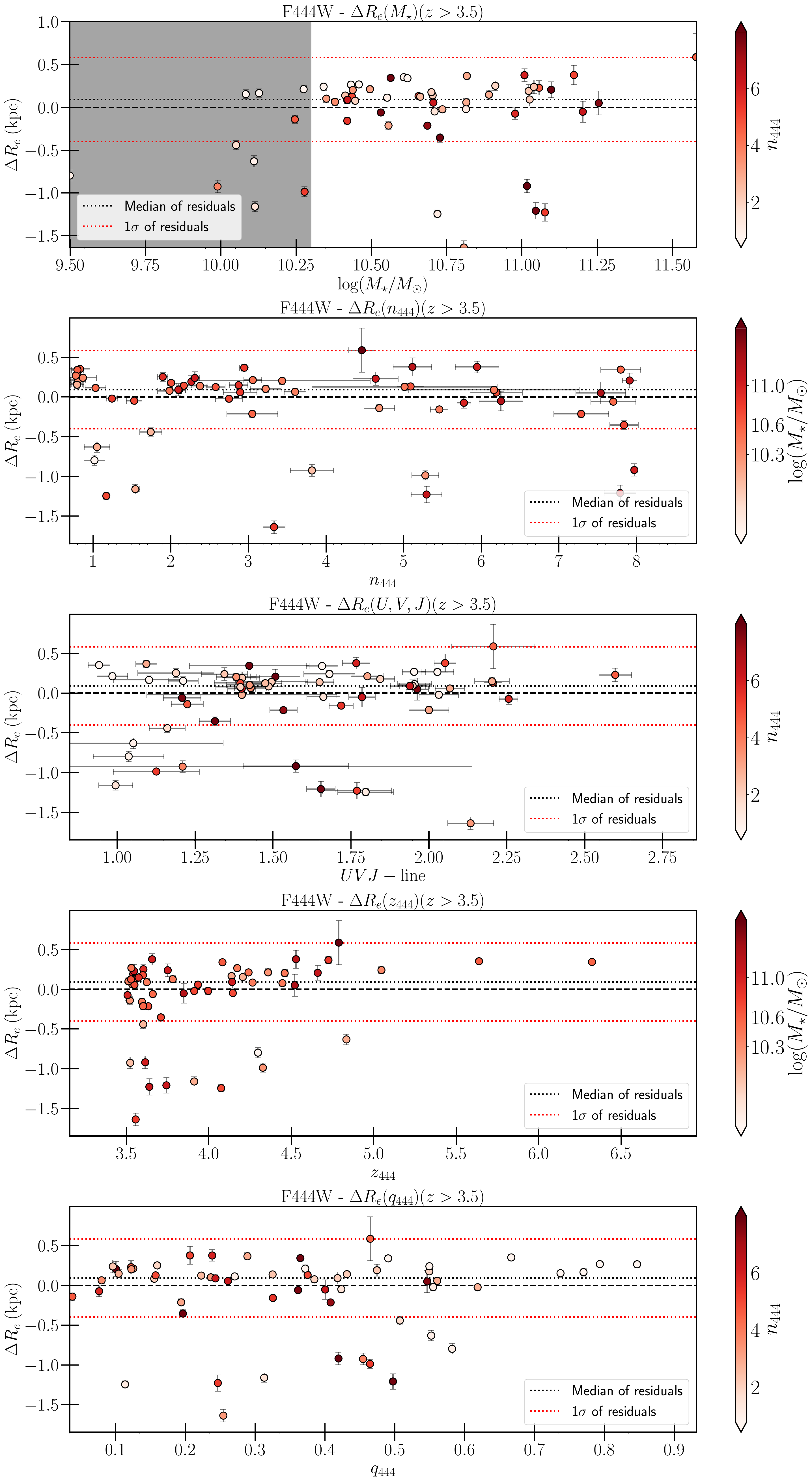}
        \caption{Residuals in the high-redshift bin of the size plotted against different parameters used in the significance study.}
        \label{fig:residuals_up}
    \end{subfigure}
    \hfill  
    \begin{subfigure}[b]{0.48\textwidth}
        \centering
        \includegraphics[width=\textwidth]{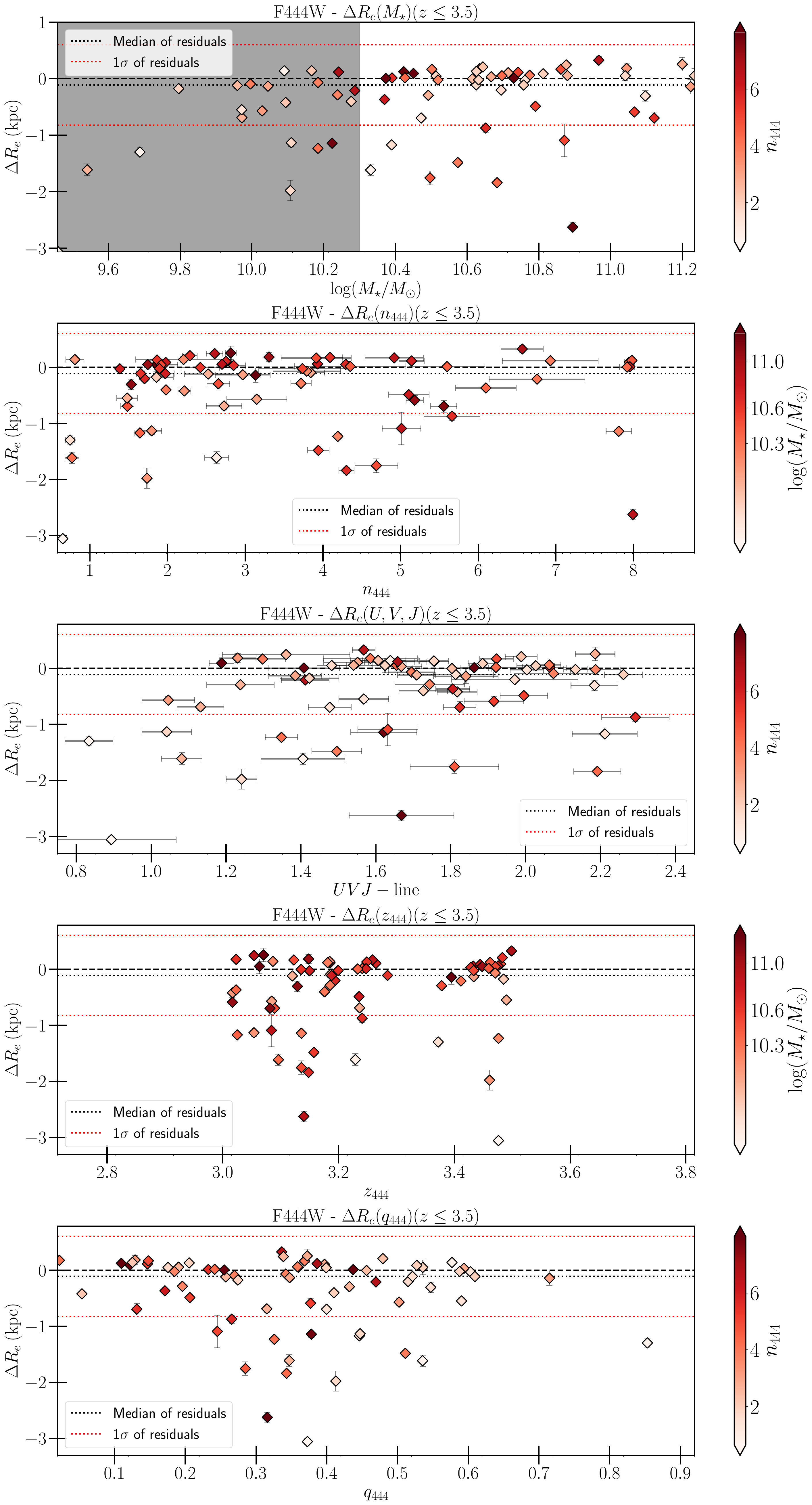}
        \caption{Residuals in the low-redshift bin of the size plotted against different parameters used in the significance study.}
        \label{fig:residuals_bot}
    \end{subfigure}
    \caption{Comparison of residuals $\Delta R (\theta) \coloneqq R_\mathrm{obs}-R_\mathrm{fit}$ in high- and low-redshift bin for F444W. The red dotted line bounds the $1\sigma$ limits of the residuals. The color-code is set by the covariate variable on the vertical bar to the right. The gray shaded area is as in Fig. \ref{fig:F200+F444_split}.}
    \label{fig:FinalForwards_app}
\end{figure}
\begin{figure*}[]
    \centering
    \includegraphics[scale=0.3]{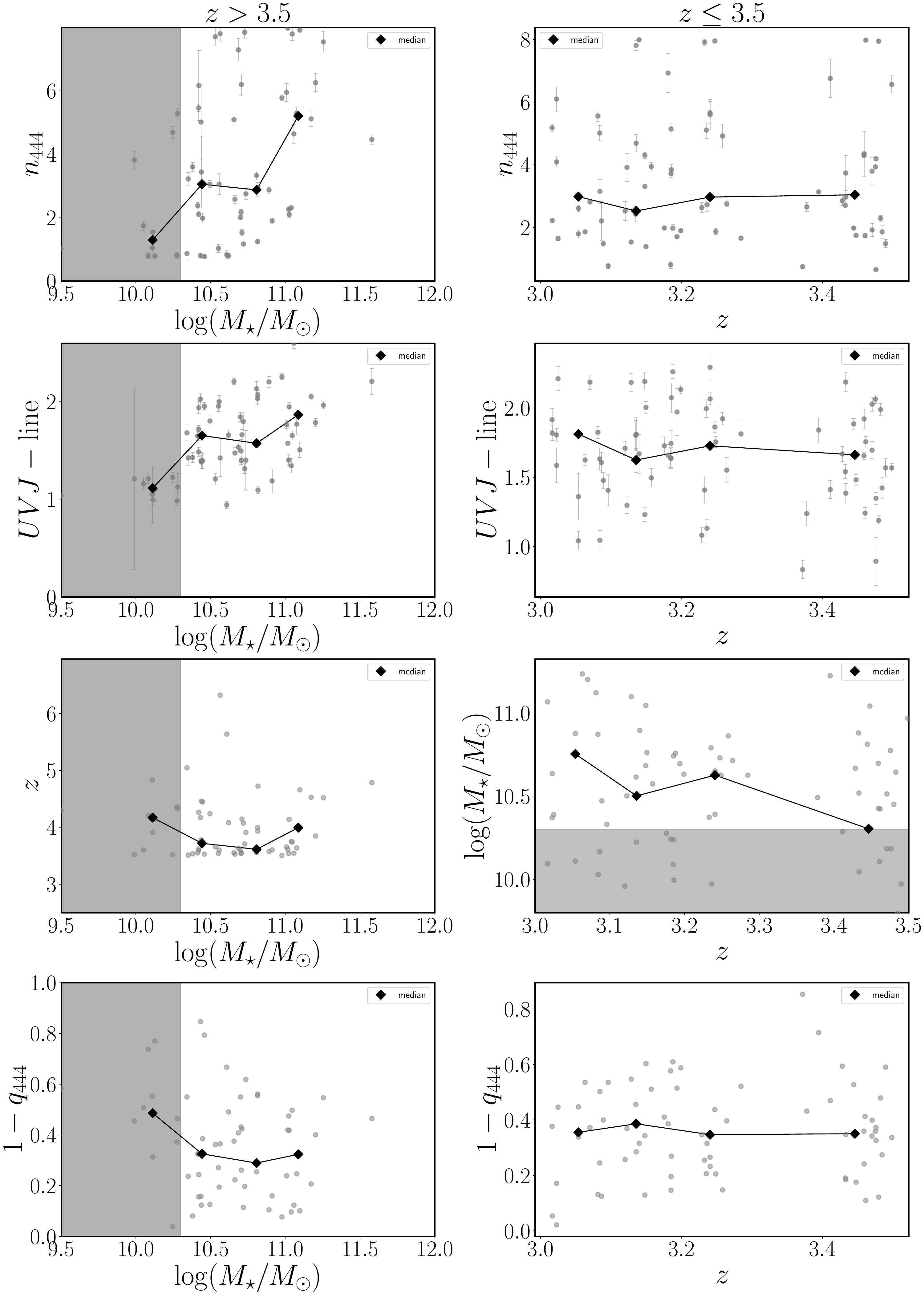}
    \caption{The filled circles are the observed points, whereas the rhombic points represent the medians in the bins of the corresponding independent variable. The gray shaded area is as in Fig. \ref{fig:F200+F444_split}.}
    \label{fig:paramtoparam}
\end{figure*}
\\ Regarding the study of the best predictor (Figure \ref{fig:FinalForwards_2}), the statistical sample hereby used is larger than previously used samples, therefore it allows a better statistical characterization of the size evolution.
\section{Table}
We present a table with the main results of the study allowing for quick consultation. A more comprehensive file will be available online with more data and figures.
\onecolumn
\begin{landscape}
    \begin{longtable}{l|lllclllll}
            \toprule
                ID\tablefootmark{a} & R.A.(deg) & Decl.(deg) & $z_\mathrm{phot}$($z_\mathrm{spec}$)\tablefootmark{b}& $\log(M_\star / M_\odot)$ & $R_\mathrm{f444w}$(kpc)\tablefootmark{c} & $R_\mathrm{f200w}$(kpc)\tablefootmark{c} & $n_\mathrm{f444w}$ & $1 - q_\mathrm{f444w}$ \\
            \midrule
            0 &  215.039058 &  53.002781 &  4.00( 4.2898) &  10.81 &  0.595 $\pm$  0.007 &  0.786 $\pm$  0.014 &  1.243 $\pm$  0.074 &  0.555 $\pm$  0.010\\
1 &  214.909553 &  52.875028 &  3.38( 3.3440) &  10.49 &  0.721 $\pm$  0.014 &  0.851 $\pm$  0.019 &  2.653 $\pm$  0.147 &  0.432 $\pm$  0.015\\
2 &  214.865249 &  52.810842 &  4.14 &  10.13 &  0.139 $\pm$  0.002 &  0.138 $\pm$  0.001 &  0.791 $\pm$  0.071 &  0.770 $\pm$  0.104\\
3 &  214.915546 &  52.949018 &  4.66( 4.9000) &  11.10 &  0.539 $\pm$  0.010 &  0.408 $\pm$  0.017 &  7.911 $\pm$  0.093 &  0.101 $\pm$  0.013\\
4 &  214.957885 &  52.980300 &  3.54 &  10.70 &  0.378 $\pm$  0.006 &  0.371 $\pm$  0.005 &  2.168 $\pm$  0.092 &  0.431 $\pm$  0.013\\
5 &  214.895613 &  52.856498 &  3.15( 3.2396) &  11.04 &  0.527 $\pm$  0.003 &  0.556 $\pm$  0.003 &  3.306 $\pm$  0.040 &  0.129 $\pm$  0.005\\
6 &  214.866053 &  52.884257 &  3.54( 3.4343) &  11.02 &  0.505 $\pm$  0.002 &  0.451 $\pm$  0.002 &  2.266 $\pm$  0.025 &  0.474 $\pm$  0.003\\
7 &  214.897046 &  52.792221 &  3.19 &  10.70 &  0.717 $\pm$  0.004 &  0.960 $\pm$  0.011 &  1.702 $\pm$  0.042 &  0.515 $\pm$  0.005\\
8 &  214.904848 &  52.935350 &  3.26 &  10.71 &  0.419 $\pm$  0.004 &  0.426 $\pm$  0.005 &  2.752 $\pm$  0.091 &  0.397 $\pm$  0.010\\
9 &  214.858988 &  52.895089 &  3.91 &  10.11 &  1.462 $\pm$  0.031 &  1.610 $\pm$  0.069 &  1.544 $\pm$  0.058 &  0.313 $\pm$  0.015\\
10 &  214.879098 &  52.888065 &  3.54( 3.4439) &  10.38 &  0.321 $\pm$  0.005 &  0.227 $\pm$  0.004 &  3.602 $\pm$  0.142 &  0.080 $\pm$  0.016\\
11 &  214.941737 &  52.884560 &  3.24( 3.2195) &  10.39 &  0.377 $\pm$  0.010 &  1.906 $\pm$  0.170 &  5.598 $\pm$  0.488 &  0.233 $\pm$  0.023\\
12 &  214.981811 &  52.991239 &  3.43 &  10.88 &  0.556 $\pm$  0.004 &  0.577 $\pm$  0.005 &  2.700 $\pm$  0.053 &  0.191 $\pm$  0.007\\
13 &  215.065856 &  52.932949 &  3.61 &  11.02 &  1.612 $\pm$  0.022 &  0.735 $\pm$  0.018 &  7.972 $\pm$  0.030 &  0.419 $\pm$  0.007\\
14 &  214.806413 &  52.742785 &  4.79 &  11.58 &  0.574 $\pm$  0.008 &  2.256 $\pm$  0.137 &  4.462 $\pm$  0.170 &  0.465 $\pm$  0.007\\
15 &  214.707437 &  52.752597 &  3.02 &  10.64 &  0.311 $\pm$  0.005 &  0.251 $\pm$  0.005 &  4.091 $\pm$  0.165 &  0.022 $\pm$  0.015\\
16 &  214.989264 &  52.847159 &  3.28 &  10.63 &  0.594 $\pm$  0.005 &  0.683 $\pm$  0.005 &  1.652 $\pm$  0.051 &  0.521 $\pm$  0.008\\
17 &  214.808171 &  52.832213 &  4.17 &  10.43 &  0.138 $\pm$  0.002 &  0.137 $\pm$  0.001 &  0.800 $\pm$  0.073 &  0.847 $\pm$  0.052\\
18 &  214.997037 &  52.983725 &  4.30 &  9.50 &  0.969 $\pm$  0.035 &  1.153 $\pm$  0.063 &  1.017 $\pm$  0.135 &  0.582 $\pm$  0.030\\
19 &  214.767252 &  52.817699 &  3.63 &  10.69 &  0.725 $\pm$  0.019 &  0.434 $\pm$  0.014 &  7.288 $\pm$  0.352 &  0.408 $\pm$  0.011\\
20 &  214.971184 &  52.854881 &  3.55 &  10.41 &  0.259 $\pm$  0.004 &  0.270 $\pm$  0.010 &  2.377 $\pm$  0.102 &  0.325 $\pm$  0.015\\
21 &  214.836845 &  52.873456 &  3.19( 3.2157) &  10.74 &  0.424 $\pm$  0.004 &  0.332 $\pm$  0.007 &  5.141 $\pm$  0.162 &  0.147 $\pm$  0.010\\
22 &  215.012178 &  52.904487 &  3.52( 3.4715) &  10.25 &  0.482 $\pm$  0.009 &  0.463 $\pm$  0.015 &  4.686 $\pm$  0.196 &  0.038 $\pm$  0.018\\
23 &  53.165989 & -27.876587 &  3.19 &  10.00 &  0.364 $\pm$  0.005 &  1.042 $\pm$  0.262 &  3.843 $\pm$  0.178 &  0.269 $\pm$  0.014\\
24 &  53.140751 & -27.873557 &  3.48 &  9.46 &  3.223 $\pm$  0.026 &  1.653 $\pm$  0.293 &  0.650 $\pm$  0.001 &  0.373 $\pm$  0.008\\
25 &  53.181242 & -27.756509 &  3.46 &  10.70 &  0.464 $\pm$  0.002 &  0.616 $\pm$  0.003 &  4.294 $\pm$  0.042 &  0.360 $\pm$  0.003\\
26 &  53.120011 & -27.852026 &  3.48( 3.5850) &  10.45 &  0.319 $\pm$  0.003 &  0.338 $\pm$  0.004 &  7.945 $\pm$  0.064 &  0.122 $\pm$  0.009\\
27 &  53.113323 & -27.869945 &  3.50 &  10.97 &  0.335 $\pm$  0.004 &  0.456 $\pm$  0.004 &  6.567 $\pm$  0.272 &  0.336 $\pm$  0.009\\
28 &  53.108213 & -27.825183 &  4.73( 4.6530) &  10.82 &  0.209 $\pm$  0.002 &  0.222 $\pm$  0.002 &  2.946 $\pm$  0.061 &  0.289 $\pm$  0.009\\
29 &  53.139478 & -27.874138 &  3.23 &  9.54 &  1.791 $\pm$  0.090 &  1.271 $\pm$  0.047 &  2.630 $\pm$  0.154 &  0.347 $\pm$  0.017\\
30 &  53.079694 & -27.838200 &  3.48 &  10.18 &  1.555 $\pm$  0.013 &  1.535 $\pm$  0.017 &  4.192 $\pm$  0.046 &  0.326 $\pm$  0.004\\
31 &  53.081880 & -27.828799 &  4.27( 4.3439) &  10.42 &  0.316 $\pm$  0.004 &  0.390 $\pm$  0.006 &  2.105 $\pm$  0.083 &  0.156 $\pm$  0.013\\
32 &  53.055877 & -27.815659 &  4.07 &  10.72 &  1.772 $\pm$  0.026 &  2.022 $\pm$  0.130 &  1.170 $\pm$  0.036 &  0.114 $\pm$  0.014\\
33 &  53.062272 & -27.875041 &  4.24( 4.4278) &  10.50 &  0.217 $\pm$  0.004 &  0.288 $\pm$  0.007 &  3.056 $\pm$  0.116 &  0.126 $\pm$  0.015\\
34 &  53.082830 & -27.866184 &  3.51 &  10.98 &  0.741 $\pm$  0.008 &  1.802 $\pm$  0.023 &  5.777 $\pm$  0.090 &  0.077 $\pm$  0.004\\
35 &  53.130479 & -27.791201 &  3.59 &  10.42 &  0.557 $\pm$  0.005 &  0.922 $\pm$  0.011 &  5.460 $\pm$  0.113 &  0.325 $\pm$  0.005\\
36 &  53.087369 & -27.840000 &  3.46( 3.4773) &  10.11 &  2.279 $\pm$  0.173 &  1.742 $\pm$  0.036 &  1.734 $\pm$  0.065 &  0.413 $\pm$  0.083\\
37 &  53.078715 & -27.839608 &  3.46( 3.4926) &  10.42 &  0.279 $\pm$  0.003 &  0.464 $\pm$  0.009 &  7.979 $\pm$  0.025 &  0.110 $\pm$  0.008\\
38 &  53.165315 & -27.814137 &  3.06( 3.0631) &  11.23 &  0.795 $\pm$  0.001 &  0.800 $\pm$  0.215 &  1.855 $\pm$  0.004 &  0.536 $\pm$  0.001\\
39 &  53.076499 & -27.864165 &  3.48 &  10.77 &  0.494 $\pm$  0.001 &  0.621 $\pm$  0.004 &  3.931 $\pm$  0.035 &  0.359 $\pm$  0.002\\
40 &  53.196914 & -27.760531 &  3.44( 3.6080) &  10.81 &  0.487 $\pm$  0.001 &  0.673 $\pm$  0.002 &  1.976 $\pm$  0.018 &  0.527 $\pm$  0.002\\
41 &  53.138030 & -27.868294 &  3.56 &  10.66 &  0.365 $\pm$  0.004 &  0.757 $\pm$  0.020 &  5.090 $\pm$  0.173 &  0.375 $\pm$  0.009\\
42 &  34.399676 & -5.136348 &  4.52( 4.6224) &  11.25 &  0.812 $\pm$  0.021 &  0.587 $\pm$  0.058 &  7.539 $\pm$  0.320 &  0.546 $\pm$  0.009\\
43 &  34.488769 & -5.104983 &  4.36 &  10.28 &  0.138 $\pm$  0.004 &  0.135 $\pm$  0.001 &  0.806 $\pm$  0.098 &  0.372 $\pm$  0.167\\
44 &  34.492506 & -5.178661 &  3.15 &  10.76 &  0.576 $\pm$  0.005 &  0.722 $\pm$  0.008 &  1.385 $\pm$  0.044 &  0.603 $\pm$  0.007\\
45 &  34.467205 & -5.089531 &  3.24 &  10.79 &  1.050 $\pm$  0.035 &  2.083 $\pm$  0.086 &  5.106 $\pm$  0.261 &  0.206 $\pm$  0.017\\
46 &  34.500496 & -5.127265 &  4.21( 2.7611) &  10.08 &  0.139 $\pm$  0.004 &  0.137 $\pm$  0.001 &  0.793 $\pm$  0.086 &  0.737 $\pm$  0.135\\
47 &  34.496118 & -5.161035 &  3.60 &  10.91 &  0.376 $\pm$  0.004 &  0.345 $\pm$  0.003 &  1.897 $\pm$  0.065 &  0.160 $\pm$  0.011\\
48 &  34.513666 & -5.157806 &  3.14 &  10.89 &  3.245 $\pm$  0.067 &  0.695 $\pm$  0.039 &  7.992 $\pm$  0.010 &  0.316 $\pm$  0.012\\
49 &  34.227642 & -5.099286 &  3.26( 3.1348) &  10.86 &  0.433 $\pm$  0.012 &  0.455 $\pm$  0.016 &  4.917 $\pm$  0.375 &  0.148 $\pm$  0.025\\
50 &  34.209839 & -5.091602 &  3.60( 3.8250) &  10.70 &  0.340 $\pm$  0.002 &  0.407 $\pm$  0.010 &  2.005 $\pm$  0.048 &  0.549 $\pm$  0.006\\
51 &  34.524550 & -5.180366 &  3.60 &  10.56 &  0.666 $\pm$  0.028 &  1.549 $\pm$  0.148 &  3.051 $\pm$  0.328 &  0.194 $\pm$  0.034\\
52 &  34.381580 & -5.092539 &  3.39 &  11.22 &  0.979 $\pm$  0.006 &  1.249 $\pm$  0.011 &  3.132 $\pm$  0.059 &  0.715 $\pm$  0.004\\
53 &  34.484757 & -5.201705 &  4.83 &  10.11 &  0.932 $\pm$  0.043 &  0.879 $\pm$  0.092 &  1.050 $\pm$  0.164 &  0.552 $\pm$  0.037\\
54 &  34.509977 & -5.189361 &  3.18 &  10.09 &  0.157 $\pm$  0.006 &  0.152 $\pm$  0.001 &  0.806 $\pm$  0.116 &  0.577 $\pm$  0.205\\
55 &  34.227007 & -5.160354 &  3.02 &  10.09 &  0.718 $\pm$  0.011 &  0.937 $\pm$  0.044 &  2.215 $\pm$  0.082 &  0.054 $\pm$  0.015\\
56 &  34.485164 & -5.157813 &  3.41( 3.7178) &  10.29 &  0.564 $\pm$  0.023 &  0.810 $\pm$  0.072 &  6.757 $\pm$  0.617 &  0.470 $\pm$  0.022\\
57 &  34.317031 & -5.127611 &  3.93( 4.0941) &  10.82 &  0.517 $\pm$  0.013 &  0.771 $\pm$  0.020 &  2.895 $\pm$  0.216 &  0.561 $\pm$  0.020\\
58 &  34.278202 & -5.114131 &  3.18 &  10.28 &  0.753 $\pm$  0.003 &  0.787 $\pm$  0.027 &  1.979 $\pm$  0.030 &  0.410 $\pm$  0.004\\
59 &  34.322553 & -5.171386 &  3.91( 3.9420) &  10.74 &  0.557 $\pm$  0.010 &  0.739 $\pm$  0.028 &  2.751 $\pm$  0.172 &  0.619 $\pm$  0.016\\
60 &  34.365084 & -5.148848 &  4.53( 4.6200) &  11.17 &  0.422 $\pm$  0.007 &  0.532 $\pm$  0.051 &  5.114 $\pm$  0.249 &  0.207 $\pm$  0.014\\
61 &  34.394857 & -5.186418 &  3.02 &  11.07 &  1.314 $\pm$  0.017 &  1.775 $\pm$  0.050 &  5.181 $\pm$  0.111 &  0.377 $\pm$  0.006\\
62 &  34.243720 & -5.184243 &  3.14 &  10.22 &  1.475 $\pm$  0.030 &  1.558 $\pm$  0.119 &  7.810 $\pm$  0.163 &  0.378 $\pm$  0.007\\
63 &  34.381077 & -5.091735 &  3.55 &  11.06 &  0.490 $\pm$  0.012 &  0.492 $\pm$  0.023 &  4.639 $\pm$  0.291 &  0.122 $\pm$  0.021\\
64 &  34.232657 & -5.101269 &  3.02 &  10.37 &  0.750 $\pm$  0.028 &  1.067 $\pm$  0.049 &  6.099 $\pm$  0.390 &  0.171 $\pm$  0.019\\
65 &  34.405765 & -5.112022 &  3.25 &  10.63 &  0.353 $\pm$  0.005 &  0.342 $\pm$  0.005 &  1.863 $\pm$  0.091 &  0.206 $\pm$  0.013\\
66 &  34.442887 & -5.113055 &  3.24 &  9.97 &  0.954 $\pm$  0.040 &  1.870 $\pm$  0.146 &  2.727 $\pm$  0.229 &  0.315 $\pm$  0.024\\
67 &  34.338214 & -5.166201 &  3.54 &  10.55 &  0.338 $\pm$  0.012 &  0.520 $\pm$  0.012 &  1.031 $\pm$  0.133 &  0.271 $\pm$  0.023\\
68 &  34.339293 & -5.166742 &  3.19 &  10.76 &  0.658 $\pm$  0.010 &  0.619 $\pm$  0.021 &  1.971 $\pm$  0.115 &  0.609 $\pm$  0.013\\
69 &  34.344928 & -5.292577 &  3.45 &  11.04 &  0.658 $\pm$  0.005 &  0.615 $\pm$  0.006 &  1.744 $\pm$  0.038 &  0.176 $\pm$  0.008\\
70 &  34.502483 & -5.326010 &  3.02 &  10.39 &  1.561 $\pm$  0.027 &  1.888 $\pm$  0.078 &  1.643 $\pm$  0.064 &  0.446 $\pm$  0.012\\
71 &  34.233627 & -5.283850 &  3.56( 3.6989) &  10.81 &  2.210 $\pm$  0.067 &  3.182 $\pm$  0.238 &  3.330 $\pm$  0.142 &  0.255 $\pm$  0.018\\
72 &  34.290463 & -5.262103 &  3.55 &  10.71 &  0.465 $\pm$  0.010 &  0.320 $\pm$  0.006 &  6.194 $\pm$  0.338 &  0.261 $\pm$  0.014\\
73 &  34.368535 & -5.299475 &  3.57( 3.9930) &  10.89 &  0.468 $\pm$  0.006 &  0.595 $\pm$  0.018 &  2.874 $\pm$  0.118 &  0.105 $\pm$  0.014\\
74 &  34.480627 & -5.319479 &  3.52 &  9.99 &  1.194 $\pm$  0.051 &  0.988 $\pm$  0.043 &  3.819 $\pm$  0.277 &  0.455 $\pm$  0.019\\
75 &  34.272098 & -5.224842 &  3.09 &  10.17 &  0.177 $\pm$  0.021 &  0.167 $\pm$  0.008 &  2.205 $\pm$  0.615 &  0.125 $\pm$  0.086\\
76 &  34.280562 & -5.217176 &  4.14 &  11.03 &  0.606 $\pm$  0.007 &  0.615 $\pm$  0.018 &  2.099 $\pm$  0.091 &  0.418 $\pm$  0.011\\
77 &  34.344110 & -5.239480 &  3.16 &  10.57 &  1.944 $\pm$  0.049 &  1.948 $\pm$  0.064 &  3.941 $\pm$  0.139 &  0.511 $\pm$  0.009\\
78 &  34.366902 & -5.284188 &  5.64 &  10.61 &  0.123 $\pm$  0.005 &  0.121 $\pm$  0.003 &  0.830 $\pm$  0.128 &  0.667 $\pm$  0.186\\
79 &  34.405391 & -5.223527 &  6.32 &  10.56 &  0.113 $\pm$  0.002 &  0.148 $\pm$  0.038 &  7.797 $\pm$  0.257 &  0.365 $\pm$  0.084\\
80 &  34.289484 & -5.269830 &  3.66 &  11.01 &  0.310 $\pm$  0.004 &  0.659 $\pm$  0.018 &  5.946 $\pm$  0.282 &  0.239 $\pm$  0.012\\
81 &  34.511974 & -5.251536 &  3.47 &  10.51 &  0.391 $\pm$  0.011 &  0.382 $\pm$  0.011 &  1.916 $\pm$  0.211 &  0.399 $\pm$  0.028\\
82 &  34.489293 & -5.298891 &  3.51 &  10.35 &  0.273 $\pm$  0.007 &  0.184 $\pm$  0.007 &  3.223 $\pm$  0.199 &  0.237 $\pm$  0.024\\
83 &  34.504817 & -5.303292 &  3.48 &  9.80 &  0.401 $\pm$  0.014 &  0.496 $\pm$  0.046 &  1.855 $\pm$  0.216 &  0.274 $\pm$  0.039\\
84 &  34.250477 & -5.256897 &  3.46 &  10.43 &  0.386 $\pm$  0.019 &  0.566 $\pm$  0.061 &  4.351 $\pm$  0.718 &  0.242 $\pm$  0.046\\
85 &  34.258909 & -5.232333 &  3.13( 3.1926) &  11.10 &  1.051 $\pm$  0.007 &  1.507 $\pm$  0.017 &  1.532 $\pm$  0.033 &  0.547 $\pm$  0.005\\
86 &  34.404697 & -5.254782 &  3.47( 3.8580) &  10.18 &  0.390 $\pm$  0.014 &  0.458 $\pm$  0.027 &  3.791 $\pm$  0.435 &  0.342 $\pm$  0.033\\
87 &  34.441050 & -5.305485 &  3.08 &  10.87 &  1.697 $\pm$  0.285 &  1.629 $\pm$  0.090 &  5.012 $\pm$  0.249 &  0.245 $\pm$  0.167\\
88 &  34.340387 & -5.241281 &  3.85 &  11.20 &  0.873 $\pm$  0.024 &  1.368 $\pm$  0.079 &  6.254 $\pm$  0.283 &  0.400 $\pm$  0.009\\
89 &  34.317906 & -5.219204 &  3.12 &  10.50 &  0.266 $\pm$  0.012 &  0.246 $\pm$  0.009 &  3.914 $\pm$  0.452 &  0.369 $\pm$  0.039\\
90 &  150.209029 &  2.349140 &  3.05( 3.0995) &  10.88 &  0.364 $\pm$  0.006 &  0.382 $\pm$  0.005 &  2.606 $\pm$  0.106 &  0.339 $\pm$  0.015\\
91 &  150.180564 &  2.398621 &  3.14 &  10.50 &  2.185 $\pm$  0.116 &  1.959 $\pm$  0.108 &  4.688 $\pm$  0.275 &  0.285 $\pm$  0.019\\
92 &  150.165537 &  2.459116 &  5.05 &  10.34 &  0.131 $\pm$  0.006 &  0.806 $\pm$  0.192 &  0.869 $\pm$  0.175 &  0.550 $\pm$  0.189\\
93 &  150.199550 &  2.351650 &  3.24 &  10.65 &  1.369 $\pm$  0.063 &  1.646 $\pm$  0.077 &  5.662 $\pm$  0.361 &  0.266 $\pm$  0.019\\
94 &  150.151805 &  2.337954 &  3.43( 3.4454) &  10.04 &  0.416 $\pm$  0.014 &  0.310 $\pm$  0.014 &  2.968 $\pm$  0.353 &  0.348 $\pm$  0.031\\
95 &  150.183893 &  2.402638 &  3.74 &  11.05 &  1.920 $\pm$  0.054 &  2.389 $\pm$  0.149 &  7.790 $\pm$  0.206 &  0.497 $\pm$  0.011\\
96 &  150.181738 &  2.239032 &  3.64 &  11.08 &  1.960 $\pm$  0.054 &  2.780 $\pm$  0.073 &  5.295 $\pm$  0.197 &  0.247 $\pm$  0.013\\
97 &  150.167654 &  2.341992 &  4.08 &  10.62 &  0.140 $\pm$  0.003 &  0.139 $\pm$  0.001 &  0.798 $\pm$  0.077 &  0.490 $\pm$  0.154\\
98 &  150.202732 &  2.287029 &  3.78 &  10.44 &  0.279 $\pm$  0.020 &  0.263 $\pm$  0.025 &  5.012 $\pm$  1.192 &  0.158 $\pm$  0.080\\
99 &  150.156488 &  2.428019 &  3.12( 3.1670) &  9.96 &  0.381 $\pm$  0.014 &  0.372 $\pm$  0.013 &  2.523 $\pm$  0.306 &  0.257 $\pm$  0.037\\
100 &  150.203701 &  2.437691 &  3.09 &  10.47 &  1.115 $\pm$  0.022 &  1.433 $\pm$  0.059 &  1.481 $\pm$  0.086 &  0.400 $\pm$  0.016\\
101 &  150.153217 &  2.184066 &  3.05 &  10.11 &  1.434 $\pm$  0.046 &  1.356 $\pm$  0.051 &  1.795 $\pm$  0.127 &  0.447 $\pm$  0.021\\
102 &  150.170709 &  2.444302 &  3.43 &  10.67 &  0.468 $\pm$  0.007 &  0.531 $\pm$  0.007 &  2.849 $\pm$  0.154 &  0.594 $\pm$  0.014\\
103 &  150.194051 &  2.372225 &  3.71 &  10.73 &  0.885 $\pm$  0.041 &  0.964 $\pm$  0.056 &  7.840 $\pm$  0.183 &  0.197 $\pm$  0.027\\
104 &  150.153453 &  2.373319 &  3.49 &  9.97 &  0.814 $\pm$  0.020 &  0.946 $\pm$  0.061 &  1.479 $\pm$  0.133 &  0.591 $\pm$  0.020\\
105 &  150.154247 &  2.341754 &  3.60 &  10.05 &  0.726 $\pm$  0.016 &  0.879 $\pm$  0.048 &  1.742 $\pm$  0.142 &  0.507 $\pm$  0.019\\
106 &  150.149092 &  2.482718 &  3.62 &  10.42 &  0.312 $\pm$  0.021 &  0.407 $\pm$  0.027 &  6.164 $\pm$  1.097 &  0.243 $\pm$  0.060\\
107 &  150.097412 &  2.242018 &  3.25( 3.2455) &  10.73 &  0.522 $\pm$  0.012 &  0.763 $\pm$  0.033 &  7.953 $\pm$  0.050 &  0.437 $\pm$  0.014\\
108 &  150.134186 &  2.322734 &  3.19 &  10.24 &  0.625 $\pm$  0.009 &  0.621 $\pm$  0.019 &  3.718 $\pm$  0.131 &  0.196 $\pm$  0.011\\
109 &  150.094941 &  2.177159 &  3.43 &  10.52 &  0.462 $\pm$  0.022 &  1.209 $\pm$  0.160 &  3.736 $\pm$  0.563 &  0.185 $\pm$  0.045\\
110 &  150.130840 &  2.413596 &  3.75 &  11.04 &  0.468 $\pm$  0.004 &  0.505 $\pm$  0.005 &  2.310 $\pm$  0.062 &  0.096 $\pm$  0.010\\
111 &  150.097330 &  2.260114 &  3.08 &  10.03 &  0.848 $\pm$  0.042 &  0.927 $\pm$  0.045 &  3.148 $\pm$  0.387 &  0.502 $\pm$  0.028\\
112 &  150.087262 &  2.396038 &  3.53( 3.7091) &  10.66 &  0.376 $\pm$  0.006 &  0.465 $\pm$  0.006 &  2.578 $\pm$  0.123 &  0.223 $\pm$  0.014\\
113 &  150.068551 &  2.342715 &  3.07 &  11.20 &  0.564 $\pm$  0.004 &  0.894 $\pm$  0.013 &  2.813 $\pm$  0.069 &  0.372 $\pm$  0.006\\
114 &  150.131640 &  2.375043 &  3.37 &  9.69 &  1.502 $\pm$  0.048 &  1.530 $\pm$  0.028 &  0.744 $\pm$  0.057 &  0.853 $\pm$  0.018\\
115 &  150.068809 &  2.338276 &  4.46 &  10.44 &  0.203 $\pm$  0.022 &  0.382 $\pm$  0.224 &  3.435 $\pm$  1.133 &  0.123 $\pm$  0.097\\
116 &  150.082837 &  2.415809 &  3.10 &  10.33 &  1.984 $\pm$  0.092 &  3.776 $\pm$  0.212 &  0.768 $\pm$  0.090 &  0.535 $\pm$  0.019\\
117 &  150.083883 &  2.223530 &  3.08 &  11.12 &  1.459 $\pm$  0.030 &  1.654 $\pm$  0.061 &  5.556 $\pm$  0.167 &  0.131 $\pm$  0.011\\
118 &  150.071435 &  2.291177 &  4.15( 4.2897) &  10.71 &  0.570 $\pm$  0.009 &  0.557 $\pm$  0.016 &  1.529 $\pm$  0.099 &  0.423 $\pm$  0.015\\
119 &  150.117779 &  2.386856 &  3.66( 3.7526) &  10.53 &  0.503 $\pm$  0.013 &  0.370 $\pm$  0.012 &  7.701 $\pm$  0.290 &  0.362 $\pm$  0.016\\
120 &  189.025768 &  62.260502 &  3.14( 3.1260) &  10.61 &  0.482 $\pm$  0.014 &  0.606 $\pm$  0.014 &  2.421 $\pm$  0.247 &  0.456 $\pm$  0.027\\
121 &  189.275447 &  62.214136 &  4.33( 4.3998) &  10.28 &  1.338 $\pm$  0.036 &  1.031 $\pm$  0.033 &  5.280 $\pm$  0.173 &  0.465 $\pm$  0.008\\
122 &  189.221459 &  62.192401 &  3.18( 3.2386) &  10.24 &  0.222 $\pm$  0.006 &  0.340 $\pm$  0.010 &  6.929 $\pm$  0.616 &  0.386 $\pm$  0.024\\
123 &  189.268084 &  62.184657 &  3.53( 3.1485) &  10.46 &  0.146 $\pm$  0.001 &  0.117 $\pm$  0.004 &  0.776 $\pm$  0.031 &  0.794 $\pm$  0.081\\
124 &  189.155785 &  62.298522 &  3.23 &  10.37 &  0.378 $\pm$  0.010 &  0.533 $\pm$  0.025 &  7.918 $\pm$  0.098 &  0.255 $\pm$  0.017\\
125 &  189.047810 &  62.278504 &  3.48 &  10.64 &  0.284 $\pm$  0.004 &  0.325 $\pm$  0.005 &  2.287 $\pm$  0.094 &  0.479 $\pm$  0.013\\
126 &  189.234669 &  62.222702 &  3.20 &  10.63 &  0.507 $\pm$  0.003 &  0.594 $\pm$  0.006 &  1.894 $\pm$  0.045 &  0.588 $\pm$  0.005\\
127 &  189.155790 &  62.298085 &  3.15 &  10.68 &  2.350 $\pm$  0.039 &  2.257 $\pm$  0.057 &  4.303 $\pm$  0.097 &  0.343 $\pm$  0.008\\
128 &  189.220160 &  62.188066 &  4.45 &  10.45 &  0.333 $\pm$  0.007 &  0.359 $\pm$  0.010 &  1.983 $\pm$  0.154 &  0.385 $\pm$  0.016\\
            \bottomrule
            \caption{List of sources of that admitted a successful \texttt{emcee} and \texttt{pysersic} fit.}
    \end{longtable}
     \label{tab:catalog}
     \tablefoot{
     \tablefoottext{a}{IDs are given as a convention. For size computation we used old versions of the catalogs for the sake of compatibility with PSF models. IDs are likely to change across versions, yet it will not be difficult to find the new ones by matching the coordinates.}
     \tablefoottext{b}{For the sources for which it is available, the spectroscopic redshift is the best redshift estimate for unique sources obtained with NIRSpec and publicly available on DJA website.}
     \tablefoottext{c}{Sizes in kpc are computed with photometric data.}
     }
\end{landscape}
\FloatBarrier 
\clearpage

\end{appendix}
\end{document}